\input harvmac
\input epsf
\noblackbox
%\draftmode
%%%
\def\IP{{\bf P}}\def\IR{{\bf R}}
\def\br{\hfill\break}\def\ni{\noindent}\def\ov#1#2{{#1 \over #2}}
\def\cx#1{{\cal #1}}\def\IP{{\bf P}}
\def\br{\hfill\break}\def\ni{\noindent}\def\ov#1#2{{{#1}\over{#2}}}

%%%

\def\IR{\bf R}

%%%
\def\us#1{\underline{#1}}
\def\nup#1({Nucl.\ Phys.\ $\us {B#1}$\ (}
\def\plt#1({Phys.\ Lett.\ $\us  {B#1}$\ (}
\def\cmp#1({Comm.\ Math.\ Phys.\ $\us  {#1}$\ (}
\def\prp#1({Phys.\ Rep.\ $\us  {#1}$\ (}
\def\prl#1({Phys.\ Rev.\ Lett.\ $\us  {#1}$\ (}
\def\prv#1({Phys.\ Rev.\ $\us  {#1}$\ (}
\def\mpl#1({Mod.\ Phys.\ Let.\ $\us  {A#1}$\ (}
\def\ijmp#1({Int.\ J.\ Mod.\ Phys.\ $\us{A#1}$\ (}
\def\tit#1|{{\it #1},\ }
%%%
\def\ovset{\supset}
\def\IP{{\bf P}}\def\IT{{T}}\def\IS{{\bf S}}\def\IR{{\bf R}}
\def\mbr{\vskip 0.5cm}
\def\p{\partial}\def\cx#1{{\cal #1}}\def\teff{\tau_{\rm eff}}
\def\nin{\noindent}
\def\ul#1{\underline{#1}}\def\rk{{\rm rk}}
\def\frac#1#2{{#1\over#2}}
\lref\sw{N. Seiberg and  E. Witten, \nup 426 (1994) 19, 
erratum: ibid {\bf 430} (1994) 396;
\nup 431 (1994) 484.}
\lref\pol{ J. Polchinski, \prl 75 (1995) 4724.}
\lref\kklmv{S. Kachru, A. Klemm, W. Lerche, P. Mayr and C. Vafa,
                \nup 459 (1996) 537.}
\lref\klmvw{A. Klemm, W. Lerche, P. Mayr, C. Vafa, N. Warner,
                \nup 477 (1996) 746.}
\lref\akv{ S. Katz, A. Klemm and C. Vafa, \nup 497 (1997) 173.}
\lref\kmv{S. Katz, P. Mayr and C. Vafa,
Adv.Theor.Math.Phys. $\us {1}$ (1998) 53.}
\lref\stro{ A. Strominger, \nup 451 (1995) 96.}
\lref\vafaiii{ M. Bershadsky, V. Sadov and C. Vafa, \nup 463  (1996) 420.}
\lref\witii{E. Witten, {\it Some comments on string dynamics},
hep-th/9507121.}
\lref\cf{P. Candelas and A. Font, \nup 511 (1998) 295.}
\lref\cani{ P. Candelas, X. C. De La Ossa, P. S. Green and L. Parkes,
\nup 359 (1991) 21.}
\lref\kv{ S. Kachru and C. Vafa, \nup 450 (1995) 69.}
\lref\feri{S. Ferrara, J. A. Harvey, A. Strominger
                  and C. Vafa, \plt 361 (1995) 59.}
\lref\syz {A. Strominger, S.-T. Yau and E. Zaslow, \nup 479 (1996) 243.}
\lref\ov{ H. Ooguri and C. Vafa, \nup 463 (1996) 55.}
\lref\kod{K. Kodaira, {\it Annals of Math.} {\bf 77} (1963) 563;
{\it Annals of Math.} {\bf 78} (1963) 1.}
\lref\duff{ M. Duff, \nup 442 (1995) 47.}
\lref\witi{ E. Witten, \nup 443 (1995) 85.}
\lref\vafai{ C. Vafa, Adv. Theor. Math. Phys. $\us{1}$ (1998) 158.}
\lref\kvii{ S. Katz and C. Vafa, \nup 497 (1997) 146.}
\lref\witm{ E. Witten, \nup 471 (1996) 195.}
\lref\seni{ A. Sen, J. High Energy Phys. $\us {9}$ (1997) 1.}
\lref\vafaii { M. Bershadsky, V. Sadov and C. Vafa, \nup 463 (1996) 398.}
\lref\lmw{ W. Lerche, P. Mayr and N. Warner, \nup 499 (1997) 125. }
\lref\dec{ B. de Wit, P. Lauwers and A. Van Proeyen, \nup 255 (1985) 269;\br
K. Becker, M. Becker and A. Strominger, unpublished.}
\lref\suo{  A.\ Klemm, W.\ Lerche, S.\ Theisen and 
               S.\ Yankielowicz,
               \plt 344 (1995) 169;\br
P. Argyres and A. Farragi, \prl  73 (1995) 3931.}
\lref\othergg{ For a list of references see e.g. \lw.}
\lref\mw{ E. J. Martinec and N. Warner, \nup 459 (1996) 97.}
\lref\lw{ W. Lerche and N. Warner, \plt 423 (1998) 79.}
\lref\witb{ E. Witten, \nup 500 (1997) 3.}
\lref\neo{ K. Hori, H. Ooguri and Y. Oz, {\it 
Strong coupling dynamics of four-dimensional N=1 gauge
                  theories from M theory five-brane}, hep-th/9706082;\br
E. Witten, \nup 507 (1997) 658. }
%%%
\Title{\vbox{
\hbox{CERN-TH/98-165}
\hbox{NSF-ITP-98-066}
\hbox{\tt hep-th/9807096}
}}
{}
\vskip-3cm
\centerline{{\titlefont Geometric Construction of}}
\vskip 0.4cm
\centerline{{\titlefont  $N=2$ 
Gauge Theories}\foot{Lecture given at the conference
{\it Quantum aspects of 
gauge theories, supersymmetry and unification}, 
Neuchatel University, 18-23 September 1997.}}
\ 
\bigskip\bigskip
\centerline{P. Mayr}
\bigskip\centerline{\it Theory Division, CERN, 1211 Geneva 23,
Switzerland}
\centerline{and}
\centerline{\it Institute for Theoretical Physics, University of California,
Santa Barbara,}
\centerline{\it CA 93106, USA}
\vskip .3in
\noindent
In this lecture we give an elementary introduction to
the natural realization of non-perturbative 
$N=2$ quantum field theories as a low energy limit
of {\it classical} string theory. We review a
systematic construction of six, five, and four
dimensional gauge theories using geometrical data,
which provides the exact, non-perturbative
solution via mirror symmetry. This construction
has lead to the exact solution of a large class of gravity free
gauge theories, including Super Yang Mills (SYM) theories as well 
as non-conventional quantum field theories without
a known Lagrangian description.
\Date{June 1998}
\newsec{Introduction: a sketch of the geometric idea}
%%%%%%%%%%%%%%
In the past few years, the understanding of non-perturbative
aspects of field and string theory has been improved drastically.
Two of the most outstanding developments have been the
exact solution of $N=2$ SYM theories by Seiberg and Witten
\sw\ and the understanding of D-branes as charged, solitonic
degrees of freedom of string theories \pol. The subject of this 
lecture deals with the relation of these two important works:
the realization, derivation and generalization of the field
theory results \sw\ from type II strings compactifications
in connection with D-brane configurations \refs{\kklmv,\klmvw,\akv,\kmv}. 
From the point of
string theory it is rather satisfying to see that the exact solution
of {\it field} theory delivers a geometric object - the Seiberg--Witten
torus $\Sigma$ - whose appearance is somehow obscure from the point
of field theory but has a very transparent meaning in terms 
of type II brane geometries. In particular, $\Sigma\times M_4$
is to be identified with the world volume of a type IIA 
five-brane \klmvw.
Thus string theory provides
a deeper understanding, and as we will see, also an improvement
and generalization of field theory aspects. 

We start with a short outline of the geometrical realization of
$N=2$ field theories in terms of type II strings.
Let us begin with a type II string compactification to 
six dimensions on a K3 manifold.
Part of the six-dimensional low energy physics will be described by 
the dimensional reduction of the ten-dimensional supergravity action. 
However there can be additional light degrees of freedom 
arising from D-branes wrapped on n-dimensional supersymmetric 
cycles $C_n$
of the compactification geometry \stro. These states are
BPS saturated, with 
masses (or more generally tensions) depending 
on the (appropriately defined) volume of the 
wrapped D-brane geometry. 

In the following we will concentrate mostly on 
supersymmetric cycles of complex dimension one,
that is projective spaces 
$\IP^1$ or, equivalently in terms of real geometry,
two-dimensional spheres $\IS^2$. For such a two-cycle $C_2$
inside the K3, the condition to be a supersymmetric cycle is
simply that it is holomorphic in one of the possible complex 
structures \vafaiii. Depending on whether we compactify type IIA
or type IIB string on this local geometry,
we obtain two different theories 
in six dimensions. This is shown in fig.1.

\vskip 0.5cm
{\baselineskip=12pt \sl
\goodbreak\midinsert
\centerline{\epsfxsize 3.2truein\epsfbox{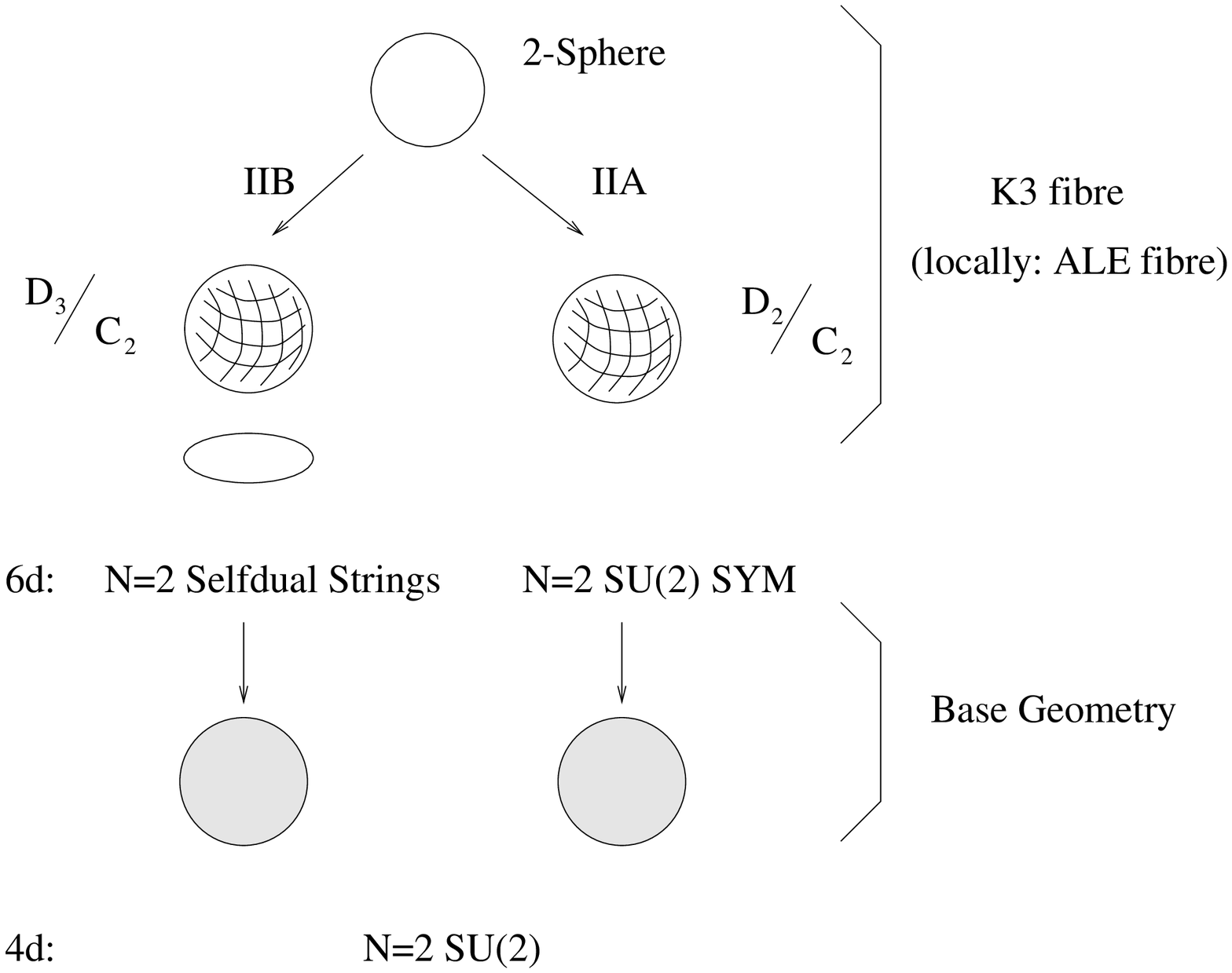}}
\leftskip 1pc\rightskip 1pc \vskip0.3cm
\noindent{\ninepoint  \baselineskip=8pt 
\centerline{{\bf Fig. 1:}
D2-brane wrappings in 6d and 4d type II compactifications.}}\endinsert}\ni

More specifically, we obtain a point-like state from the D2-brane
wrapping of the type IIA string theory. This kind of D-brane 
geometry gives rise to a pure $SU(2)$ SYM theory in six dimensions,
for reasons which will become clear later on. In the type IIB theory
we get a one-dimensional
object, that is a string, from wrapping a D3-brane \witii.
This non-critical string is of course not the same as the fundamental string we
started with. The mass/tension of the particle/string is proportional to
the volume of the two-cycle $C_2$. 

Although we get very different theories in six dimensions from 
compactification of type IIA vs. type IIB on the same geometry, 
there is a new relation on further compactification to four 
dimensions. In particular, in the compactification on a two-torus 
$\IT^2$ with $N=4$ supersymmetry,
the type IIA and type IIB theories are related by T-duality
acting on the extra torus. A similar relation holds for more
general compactification geometries with $N=2$ supersymmetry,
which are not simply products
of a K3 compactification to six dimensions and a torus compactification.
In the absence of adjoint matter representations, the geometry that
replaces the $\IT^2$ of the compactification from six to four
dimensions is again a 
$\IP^1$ and the Calabi--Yau condition requires
the total manifold to have a non-trivial fibration structure
rather than being a simple product. This means that the K3 fiber $X_2$,
whose two-cycles carry the wrapped D-brane states, varies 
holomorphically over the points 
on a new $\IP^1$, $X_2=X_2(z)$, where $z$ denotes the parameter of the
$\IP^1$.
The new $\IP^1$, which more generally can be replaced by 
a collection of intersecting $\IP^1$'s,
is called the {\it base} of the fibration.
The total space of the K3 fiber together with the base 
builds up the Calabi--Yau threefold $X_3$ on which the type IIA theory
is compactified to four dimensions.
The relation, which identifies this four-dimensional type IIA compactification
with a 
type IIB compactification on a {\it different} Calabi--Yau three-fold $X_3^*$ 
is called mirror symmetry and plays the key role in the exact solution of
the $N=2$ theory obtained in this way.\mbr

\noindent
{\it Interactions from intersections:}
To construct more interesting kind of theories with various
kinds of gauge groups and/or matter content we need
interacting D-brane states. This corresponds to a compactification
geometry with intersecting two-spheres: 
\noindent
\vskip 0.5cm
{\baselineskip=12pt \sl
\goodbreak\midinsert
\centerline{\epsfxsize 1.1truein\epsfbox{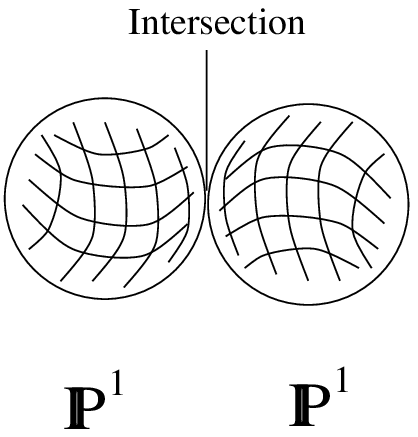}}
\leftskip 1pc\rightskip 1pc \vskip0.3cm
\noindent{\ninepoint  \baselineskip=8pt 
\centerline{{\bf Fig. 2:}
Interactions from intersecting two-spheres.
}}\endinsert}\ni
{\it Dynkin diagrams}:
Instead of drawing pictures of intersecting two-cycles as in 
fig.2, 
there is a much more convenient representation of the
type IIA D-brane geometry  which makes at the same time apparent the
amazingly close relation to group theory:
if we draw a node for each $\IP^1$ and a link for each intersection
(possibly weighted by an integer number representing multiple
intersections), we get from fig.2
the diagram shown in fig.3:
\noindent
\vskip 0.5cm
{\baselineskip=12pt \sl
\goodbreak\midinsert
\centerline{\epsfxsize 1.1truein\epsfbox{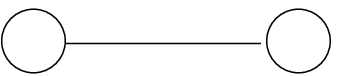}}
\leftskip 1pc\rightskip 1pc \vskip0.3cm
\noindent{\ninepoint  \baselineskip=8pt 
\centerline{{\bf Fig. 3:}
Dynkin diagram for the geometry in fig.2.
}}\endinsert}\ni
This is the Dynkin diagram of $A_2$. This is no coincidence: 
in fact the type IIA D-brane geometry in fig.2 gives rise to
a $SU(3)$ gauge system in six dimensions. This relation
between the geometry of two-cycles and the gauge group of the
type IIA theory in six dimensions will hold for all simply
laced groups, that is $A_n$, $D_n$ and $E_n$.
\mbr

\noindent
{\it Matter and product gauge groups}:
For reasons that will become clear in a moment, the pure
gauge theories with ADE gauge groups are the
only possibilities in six dimensions on purely geometrical grounds
(assuming the absence of RR background fields). 
However if we compactify further to
four dimensions, we have in addition the possibility to add 
matter, possibly charged under more than one gauge group.
As explained later on, matter arises from an extra 
localized $\IP^1$ over a point of the base $\IP^1$ as shown in 
fig.4.
\vskip 0.5cm
{\baselineskip=12pt \sl
\goodbreak\midinsert
\centerline{\epsfxsize 3.1truein\epsfbox{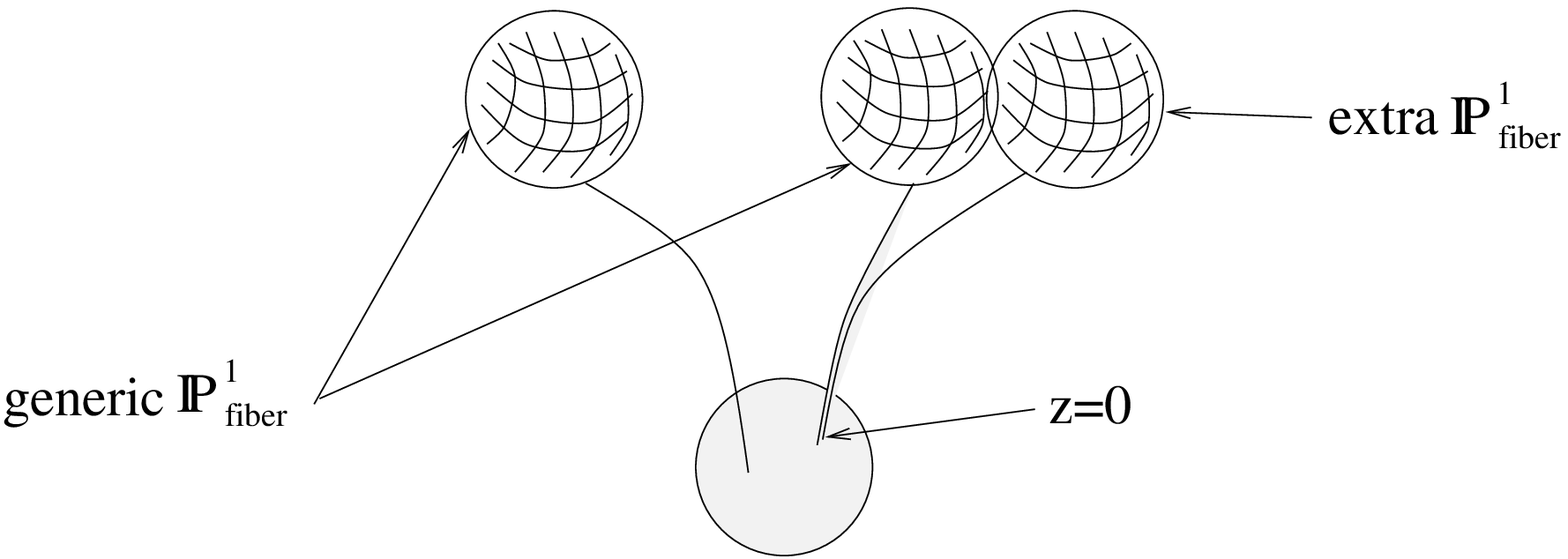}}
\leftskip 1pc\rightskip 1pc \vskip0.3cm
\noindent{\ninepoint  \baselineskip=8pt \centerline{{\bf Fig. 4:}
Matter from localized enhancement of the  singularity.
}}\endinsert}\ni
{\it More general configurations}:
The relation between geometry of intersecting two-cycles and 
Dynkin diagrams leads to the natural question of what does
it mean if we have much more general configurations of intersecting
two-cycles:
\vskip 0.5cm
{\baselineskip=12pt \sl
\goodbreak\midinsert
\centerline{\epsfxsize 4truein\epsfbox{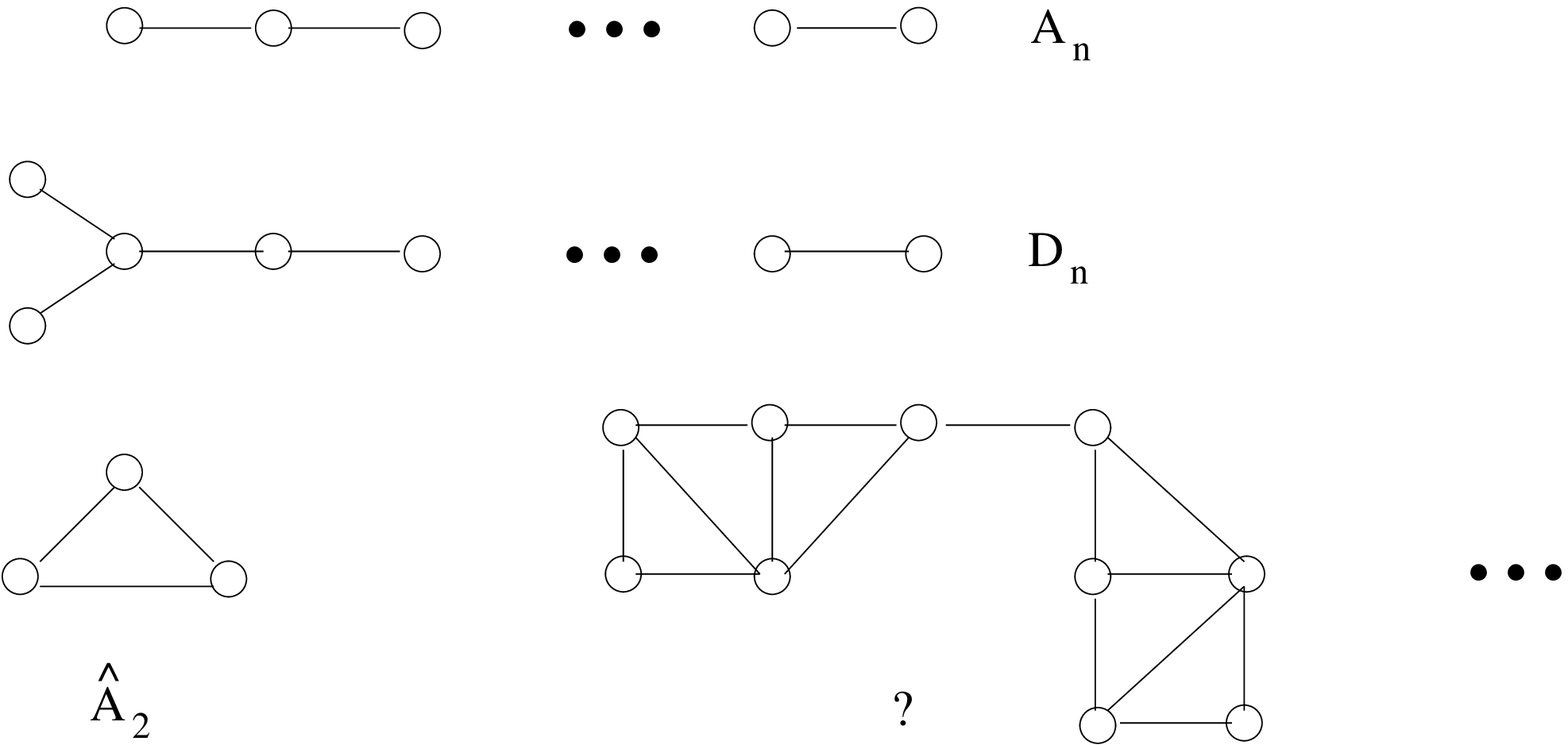}}
\leftskip 1pc\rightskip 1pc \vskip0.3cm
\noindent{\ninepoint  \baselineskip=8pt \centerline{{\bf Fig. 5:}
Diagrammatic representation of more general two-cycle configurations.
}}\endinsert}\ni
The general answer to this question is not known. The fact that the
two-cycles come as part of a Calabi--Yau geometry reduces this question
to the analysis of type IIA compactifications on general singularities 
of Calabi--Yau three-folds\foot{The fact that we have to consider {\it singular} Calabi--Yau spaces is
related to the decoupling of gravity as will be explained below.}. 
However there is a very special subclass
of $N=2$ theories arising from type IIA compactifications which we will
consider in the following: 

\item{o}        
We will restrict to a class of Calabi--Yau geometries, which
generalize K3 fibrations and can therefore be interpreted as a six-dimensional
compactification followed by a further position dependent compactification
to four dimensions. The generalization is in the following sense:
instead of considering a global K3 geometry, we consider only a local
neighborhood of the geometry of intersecting two-cycles. These geometries
are described by non-compact ALE spaces with ADE type singularities
at the origin. The two-cycles can be understood as the blow up spheres
of the resolution of the ADE singularity. The total space will be
therefore a non-compact Calabi--Yau threefold of the form of
a two complex dimensional ALE space fibered over a 
one-dimensional base geometry, which is
itself a collection of intersecting two-cycles.

\item{o}
Furthermore we consider geometries leading to $N=2$ theories which
can be consistently decoupled from gravity. This conditions restricts
the possible base geometries as well as the possible kinds of fibrations.

\nin
The second condition 
does not necessarily mean that these theories will be conventional 
$N=2$ gauge theories in four dimensions, however. Different kind
of geometries may also give rise to quantum theories without a
(known) Lagrangian formulation, such as interacting conformal 
field theories or theories involving non-critical strings.

\mbr
Of course we are not only interested in constructing these theories
as type IIA D-brane configurations, but also to solve them exactly.
Amazingly, this exact solution is immediately obtained using a 
{\it classical} symmetry of string theory, namely mirror symmetry!
 
\mbr
The present subject is related to the topics presented
in Philip Candelas lecture at this conference on 
gauge symmetries from toric polyhedra in the
context of F-theory/heterotic string duality
\ref\hulltown{C. Hull and P. Townsend, \nup 438 (1995) 109.}\kv\feri.
The type IIA D-brane
configuration that we discuss
provides the microscopic explanation for the observations
on the relation between toric polyhedra and the gauge groups of the
dual heterotic theory \cf. However note that we do not need any 
non-perturbative string duality (and in particular no heterotic
description) but only classical type II string theory for the
understanding of the gauge theory. Moreover the intrinsic objects
are D2-branes wrapped on configurations of intersecting two-cycles,
no matter how we realize this geometry; in particular this is also true for
geometries which cannot be represented in toric geometry.  
For previous lectures on the subject see 
\ref\lecs{W. Lerche,
Nucl. Phys. Proc. Suppl. $\us{B55}$ (1997) 83;\br
A. Klemm, hep-th/9705131.}.

\newsec{Basic Concepts}
\subsec{From exact $N=2$ SYM theory to string theory}
In 1994, Seiberg and Witten achieved to determine the 
exact effective action up to two derivatives of 
$N=2$ $SU(2)$ SYM theory \sw. More specifically, the
$N=2$ theory has generically Higgs branches as well as 
Coulomb branches. The Higgs branch is parametrized by
scalar fields in hypermultiplets with
flat directions and can be computed using the classical 
Lagrangian of gauge theory. On the other hand the Coulomb branch
is parametrized by the scalar fields in vector multiplets
and the exact effective action is affected by corrections from 
non-perturbative point-like instantons.
The main physical object on the Coulomb branch is 
the effective gauge coupling $\tau_{\rm eff}=\theta/\pi+8\pi i/g^2$ 
depending on the modulus on the $U(1)$ Coulomb branch, namely
the scalar component $a$ of the neutral part of the 
$SU(2)$ vector multiplet $\phi$. $\tau_{\rm eff}$ appears as
the second derivative of a holomorphic prepotential
$\cx F (a)$:
\eqn\sws{
\tau_{\rm eff}(a)=\frac{\p^2}{\p_a^2} \cx F(a) = \frac{\p_u a_D}{\p_u a} \ .
}
In the second expression, 
$a$ and $a_D$ denote the two period integrals of a certain meromorphic 
one form $\lambda$ on a complex torus $\Sigma$:
\eqn\swsii{
a=\int_{\alpha_1} \lambda,\qquad a_D=\int_{\alpha_2} \lambda \ ,
}
where $\alpha_i,\ i=1,2$ are a basis of one-cycles on $\Sigma$.
Moreover, 
$u$ is the Weyl invariant modulus $u=tr \phi^2$ parameterizing the 
complex structure of the torus. There is
an important formula for the mass of a BPS state with electric/magnetic
quantum numbers $(q_e,q_m)$ in terms of the
periods $a$ and $a_D$:
\eqn\masses{
m(q_e,q_m)=\sqrt{2}|q_e\cdot a + q_m \cdot a_D| \ .}
\mbr
Let us recall the logic of the approach of ref.\sw.
Holomorphicity of the $N=2$ gauge coupling ensures that the
exact non-perturbative gauge coupling $\teff$ is determined
by a finite set of data, namely the singularities in the moduli
space parameterized by the scalar vev together with the
local behavior at these singularities. For asymptotic free
theories, the local behavior for large values of the Coulomb parameter
is known from the perturbative spectrum. Imposing positivity of
the gauge coupling, it was possible to collect sufficiently
enough information about the extra singularities at strong 
coupling to determine the exact solution $\teff(a)$, including
all non-perturbative instanton corrections: the mathematical answer
to the problem is that $\teff$ is the period matrix of the 
torus $\Sigma$, as described by eqs.\sws,\swsii.

\vbox{
One of ones first thoughts concerning this result is:
"is there life on the torus" ? Does it have a physical 
meaning apart from its mathematical usefulness ? Is there
a realization of other physical quantities of the
$N=2$ field theory in terms of the torus ?  The answer
to this question has been found soon to be "yes" \kklmv, with
a result which appears to be 
quite a strong hint in favor of string theory:
\vskip 0.5cm
{\baselineskip=12pt \sl
\goodbreak\midinsert
\centerline{\epsfxsize 4.2truein\epsfbox{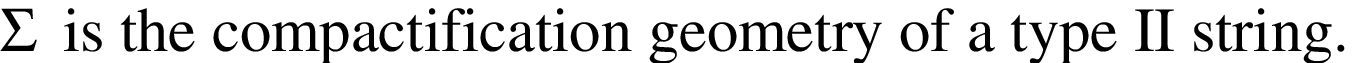}}
\leftskip 1pc\rightskip 1pc 
\vskip0.1cm
%\noindent{\ninepoint  \baselineskip=8pt {\bf Fig. 1:}}
\endinsert}\ni}
We will give a much more precise statement of this relation,
including the generalization of $\Sigma$ to other gauge groups 
later on; in particular the mathematical answer to field
theory that replaces the torus of $SU(2)$ in the case of more 
complicated gauge theories are {\it three complex dimensional
Calabi--Yau manifolds}, precisely as expected from string theory.

Moreover the quantum effects of the $N=2$ gauge theory are {\it classical}
effects from the point of string theory! This correspondence
maps instantons of the gauge theory to geometrical objects
of a type II compactification, which can be done \cani\ thanks
to the power of mirror symmetry, a symmetry of 
classical type II string theory \ref\misy{See e.g. S. T. Yau (ed.),
{\it Essays on Mirror Manifolds}, International Press 1992.}.

Historically, the use of mirror symmetry for the calculation
of space-time instanton effects has been started in the context of
type IIA/heterotic duality \refs{\kv,\feri,\kklmv}. However it is important
to note that we do not need the non-perturbative, heterotic
picture: all we need is classical type II string \
theory including the charged RR states from D-brane wrappings 
\stro.

Let us mention some advantages of the string understanding of
the torus $\Sigma$ as compared to the field theory point of view.
Firstly, we have a concrete physical meaning for the surprising
appearance of $\Sigma$ in the exact solution of the  SYM theory.
Secondly, string theory provides the framework to define additional
quantities of the SYM theory starting from the torus: $\Sigma$ appears
as part of the target geometry of the string sigma model and there is a
well-defined framework to calculate corrections, such as higher 
derivative terms or gravitational corrections. Note that 
the exact string theory solution obtained from mirror symmetry 
already includes all gravitational corrections. In fact the
decoupling of gravity is one of the non-trivial steps to obtain 
the exact solution of globally supersymmetric SYM theory \kklmv.

Moreover, string theory provides also new insights to 
field theory itself, which will be discussed in the following:
the representation of BPS states as windings of self-dual strings
on $\Sigma$ and the determination of the {\it stable} BPS spectrum
using this picture \klmvw, the systematic
generation of many new solutions with arbitrary gauge groups
from the
classification of geometric singularities, the S-duality groups
of these theories and a physical interpretation of this symmetry
\kmv.

\subsec{Three higher-dimensional string theory embeddings of $N=2$ SYM}

There are three T-dual string theory embeddings of four-dimensional
$N=2$ SYM which we will discuss in this lecture; all of them 
are related by some kind of T-duality (fig.6).
\vskip 0.5cm
{\baselineskip=12pt \sl
\goodbreak\midinsert
\centerline{\epsfxsize 4.8truein\epsfbox{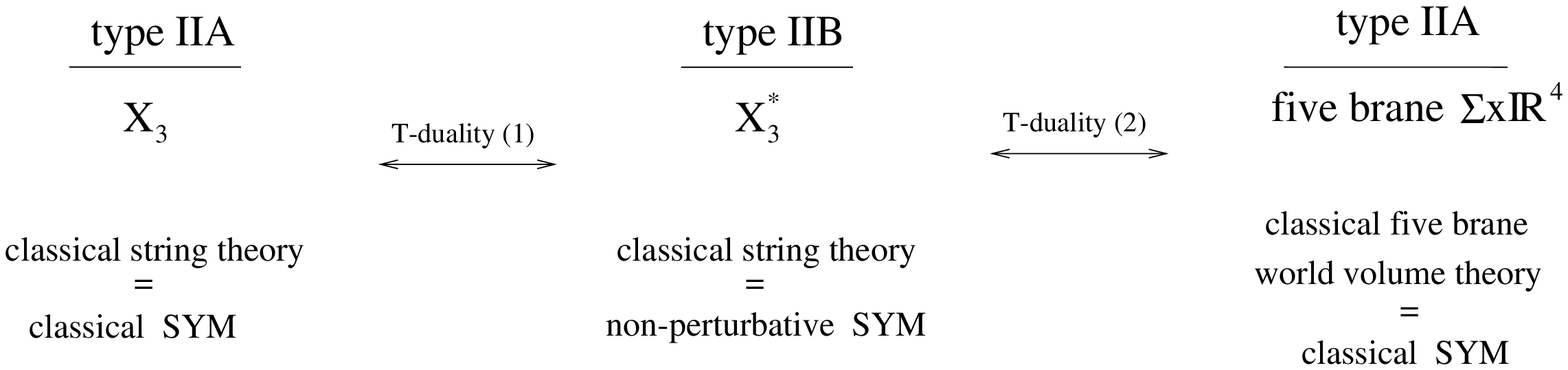}}
\leftskip 1pc\rightskip 1pc \vskip0.3cm
\noindent{\ninepoint  \baselineskip=8pt \centerline{{\bf Fig. 6:}
Three T-dual type II compactifications.
}}\endinsert}\ni
The starting point of the geometric construction is  type IIA theory
on a Calabi--Yau three-fold $X_3$. The geometry of $X_3$ contains a local
patch with a collection of intersecting two-cycles that support 
the light states which are relevant for the 
SYM theory in an appropriate region of the moduli space.
The wrapped D2-brane states together
with the massless fundamental string excitations provide the perturbative
degrees of freedom of the four-dimensional
gauge theory. The classical\foot{Both, in the space-time as well as
string worldsheet sense.} string theory 
answer agrees with the classical gauge theory answer after decoupling 
gravity. 

Mirror symmetry maps the type IIA theory on $X_3$ to a type IIB
theory on the mirror manifold $X_3^*$. This symmetry has been
interpreted in \syz\ as a T-duality transformation; for the
special geometries that we will consider the relation to T-duality
will be quite explicit. The important point for the solution
of the perturbative theory constructed in the first step is that
the classical string theory answer for type IIB on $X^*_3$ is
already the full exact result.

A third representation of the same theory is obtained from this
theory by a different T-duality transformation, which maps
type IIB on the $A_n$ singularity 
to type IIA on $n$ symmetric five-branes \ov.
The $N=2$ SYM theory appears as the world volume
theory of the five-brane, which has the geometry $\Sigma \times \IR^4$,
with $\Sigma$ the ``Seiberg-Witten geometry'' as before. We will 
give a short discussion of this representation in sect.2.9. 

What the three representations have in common is that the 
charged states of the SYM theory are represented by D-brane
states. The perturbative definition of the $N=2$ SYM theory,
determined by the root lattice corresponding to a gauge group $G$ and 
a weight lattice of matter representations
$R_i(G)$ of $G$, is in one-to-one correspondence with the D-brane 
geometry determined by the homology lattice
of two-cycles. This close link between the perturbative
definition of a $N=2$ theory and the geometrical data offers an 
interesting approach to study a large class of quantum field theories:
as we will see in a moment, the relevant geometries of two-cycles
are a special kind of singularities which are well studied 
mathematically.
\vskip 0.5cm
{\baselineskip=12pt \sl
\goodbreak\midinsert
\centerline{\epsfxsize 4.8truein\epsfbox{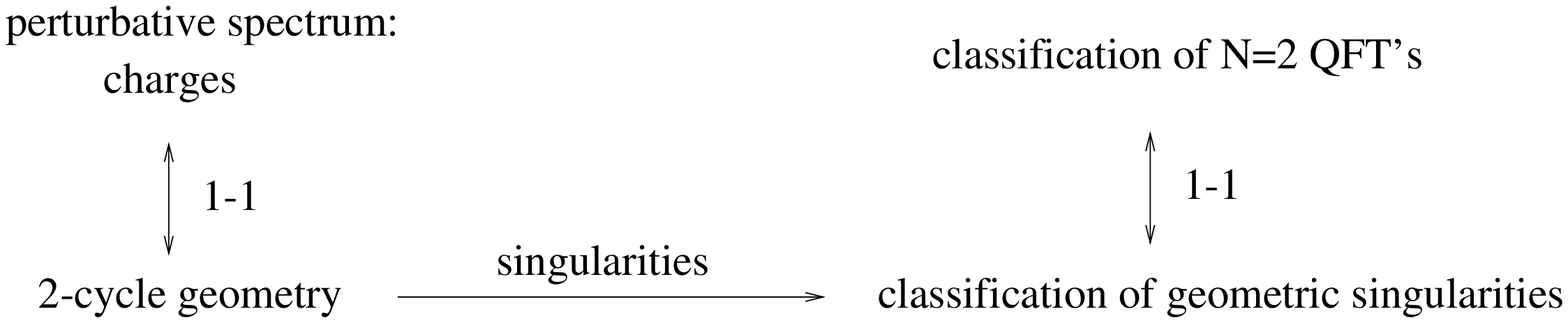}}
\leftskip 1pc\rightskip 1pc \vskip0.1cm
%\noindent{\ninepoint  \baselineskip=8pt {\bf Fig. 1:}}
\endinsert}\ni
\subsec{The starting point: $N=2$ in six dimensions}
The four-dimensional theories which we will consider can 
be understood as certain compactifications of six-dimensional
$N=2$ theories. It is useful to keep in mind this distinction
between the compactification to six dimensions followed by a further
compactification to four dimensions in this construction: 
the former will determine the gauge group $G$ of the four-dimensional
theory, whereas the second step will contain the information
about the  matter representations $R_i(G)$.

$N=2$ theories in six dimensions arise from type IIA
compactification on a K3 manifold $X_2$ or a non-compact geometry
with the same holonomy properties. The relevant data
of the K3 geometry is the homology $H_2(X_2)$ of two-cycles, which is
also dual to the cohomology of 2-forms $H^2(X_2)$. There are
two sources of particle states in the six-dimensional theory:
uncharged fields arise by dimensional reduction of the
ten-dimensional RR 3-form $A^{(3)}$:
\eqn\xxx{
A^{(3)}\rightarrow A^a \wedge \omega^a,
}
where $\omega^a \in H^2(X_2)$ is a basis of 2-forms and
$A^a$ corresponds to a neutral four-dimensional vector multiplet.
The charged fields arise from D2-branes wrapped on the
two-cycle $C^a_2$ dual to $\omega^a$; the charge arises from
the worldsheet coupling
\eqn\coupling{
\int_{C_2^a} d^3\sigma A^{(3)} \rightarrow \int d\tau A^a\ .
}
As an example consider the simple geometry with one two-cycle
$C_2$ in fig.7. It
gives rise to a $SU(2)$ gauge theory in six dimensions:
\vskip 0.5cm
{\baselineskip=12pt \sl
\goodbreak\midinsert
\centerline{\epsfxsize 1.1truein\epsfbox{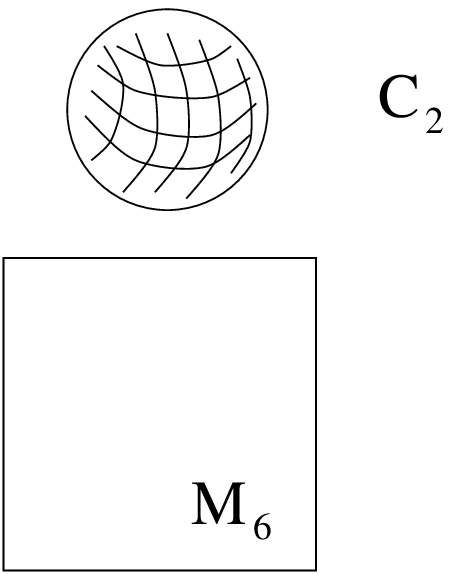}}
\leftskip 1pc\rightskip 1pc \vskip0.3cm
\noindent{\ninepoint  \baselineskip=8pt \centerline{{\bf Fig. 7:}
Geometry for the six-dimensional $SU(2)$ gauge theory.
}}\endinsert}\ni
The bosonic components of the $Z$ vector multiplet are the
vector field $A_\mu$ from the 2-form dual to $C_2$ and the
scalar component $a$ which measures the volume of $C_2$ as
defined by the K\"ahler form. The $W^\pm$
vector multiplet arises from the D2-brane wrapped on $C_2$
with the two possible orientations. The mass of these
vector bosons is proportional to the volume of $C_2$, that
is proportional to the Coulomb parameter $a$, in agreement
with field theory.

More generally, the single two-cycle $C_2$ is replaced by a 
collection of intersecting two-cycles $C^a_2$ contained in
a local piece of the K3 manifold (fig.8).
\vskip 0.5cm
{\baselineskip=12pt \sl
\goodbreak\midinsert
\centerline{\epsfxsize 3truein\epsfbox{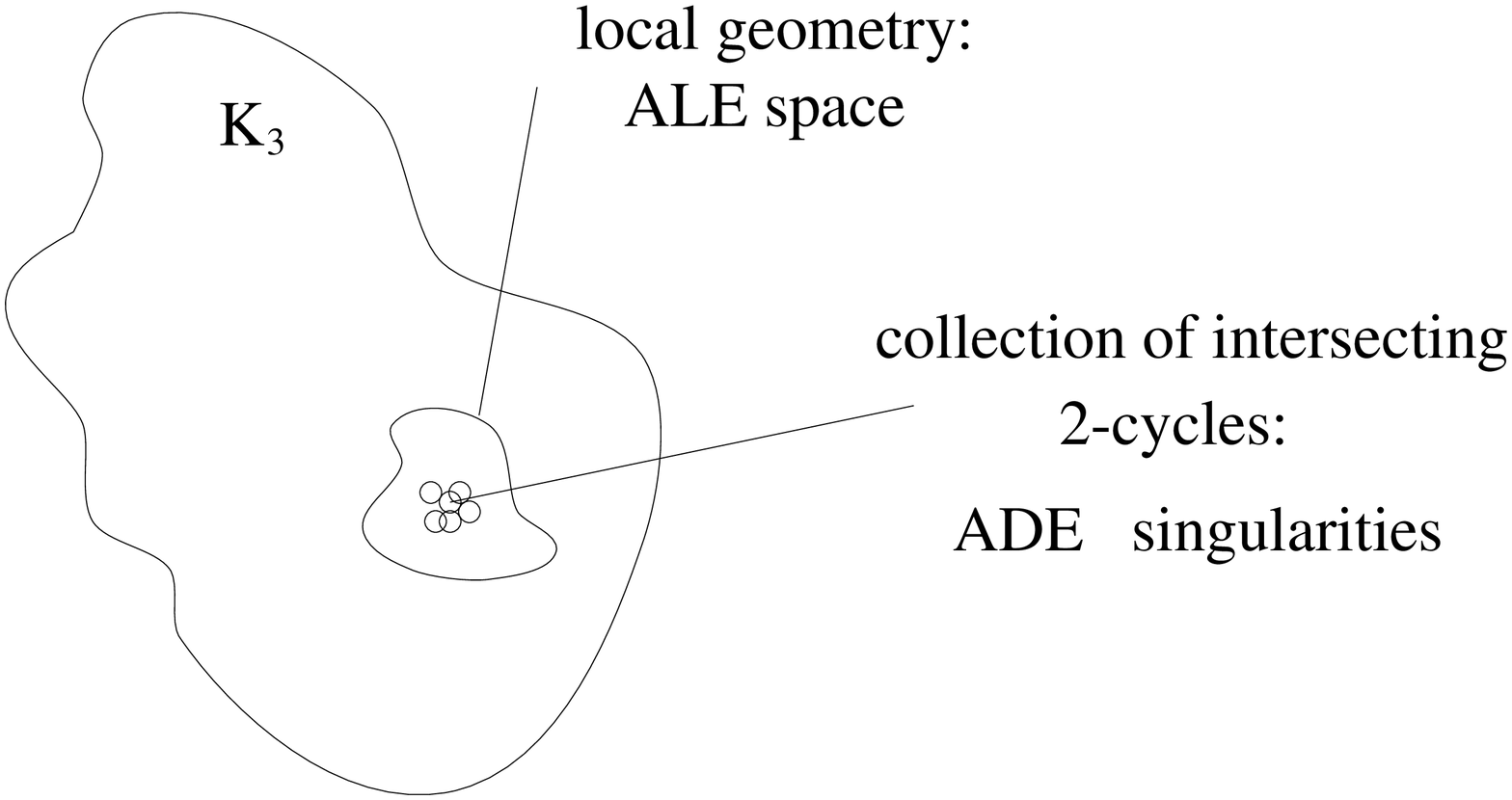}}
\leftskip 1pc\rightskip 1pc \vskip0.3cm
\noindent{\ninepoint  \baselineskip=8pt \centerline{{\bf Fig. 8:}
Local geometry of K3 singularities.
}}\endinsert}\ni
Let us first explain why we
have to consider {\it local singularities}. 
Since we want to decouple gravity, we
have to take a limit where $m_{pl}\to \infty$. However, at the same time,
we want to keep
finite the masses of the wrapped D2-brane states, e.g. the $W^\pm$ bosons,
which are proportional to the volume of the two-cycle, $m_{W^\pm}\sim
Vol(C_2)$ in units of $m_{pl}$. In other words, we have to consider 
{\it very small volumes} for the two-cycles which in turn means
to consider some sort of singular geometries.
Luckily, all singularities of a polarized K3 manifold at finite distance in the
moduli space are well-known. The homology of small two-cycles consists of
collections of $\IP^1$'s which intersect according to the Dynkin
diagrams of the simply laced groups A,D and E \ref\arnold{See e.g.
 V. Arnold , {\it Dynamical Systems VI}, Singularity Theory I,
Enc. of Math. Sci. Vol. 6, Springer 1993.}. These singularities
are called the ADE singularities of K3\foot{For recent results on
non-simply laced groups from K3 manifolds see 
\ref\chl{J.H. Schwarz and A. Ashoke Sen, \plt 357 (1995) 323;\br
M. Bershadsky, T. Pantev and V. Sadov, 
{\it F-theory with quantized fluxes}, hep-th/9805056;\br
P. Berglund, A. Klemm, P. Mayr and S. Theisen, {\it On type IIB vacua with
varying coupling constant}, hep-th/9805189.}.}

The reason that we have to consider only the {\it local} geometry of the
singularity is then obvious for the same reason: all other homology
cycles of the global geometry stay at a generic volume and give rise
to corrections from super-massive states which vanish in the limit
$m_{pl}\to \infty$. As mentioned already, 
the local geometry of the K3 ADE singularities is captured
by the so-called ALE spaces of ADE type. 
These are the geometries we have to consider in the following.

There is also a meaning to the extended Dynkin diagrams corresponding
to the affine versions of A,D and E: if the ADE singularity arises
from the collision of singular fibers of an elliptic fibrations
as classified by Kodaira \kod, there is an additional two-cycle class
(the class of a generic fiber) which corresponds to the extended
node of the Dynkin diagram. These extended Dynkin diagrams
play a very special role for the superconformal four-dimensional theories
considered in \kmv.
However, shrinking this extra two-cycle is at an infinite 
distance of the K3 moduli space and is therefore not relevant for the 
six-dimensional gauge groups. 

%If we put the type IIA string on such a local geometry, the
%homology lattice of intersecting two-cycles comes to physical
%life as the charge lattice of the D-brane states. 
%The intersections determine the interactions of the D-brane 
%states. In particular, since each two-cycle class corresponds to a 
%certain $U(1)$ gauge field from the dimensional reduction of
%the ten-dimensional three-form, intersections allow for the existence 
%of D2-brane wrappings with multiple $U(1)$ charges:

%\begin{figure}[ht]
%\vspace{0.2cm}
%\hbox to\hsize{\hss
%\epsfysize=2.3cm
%\epsffile{fig12.eps}\hss}
%\vspace{0cm}
%\caption{Multiply charged D-brane state.}
%\be\label{multi_charge}$$
%\end{figure}

\subsec{Four-dimensional theories from compactification}
To obtain four-dimensional theories we consider a further
compactification on a one complex dimensional base:
\vskip 0.5cm
{\baselineskip=12pt \sl
\goodbreak\midinsert
\centerline{\epsfxsize 2truein\epsfbox{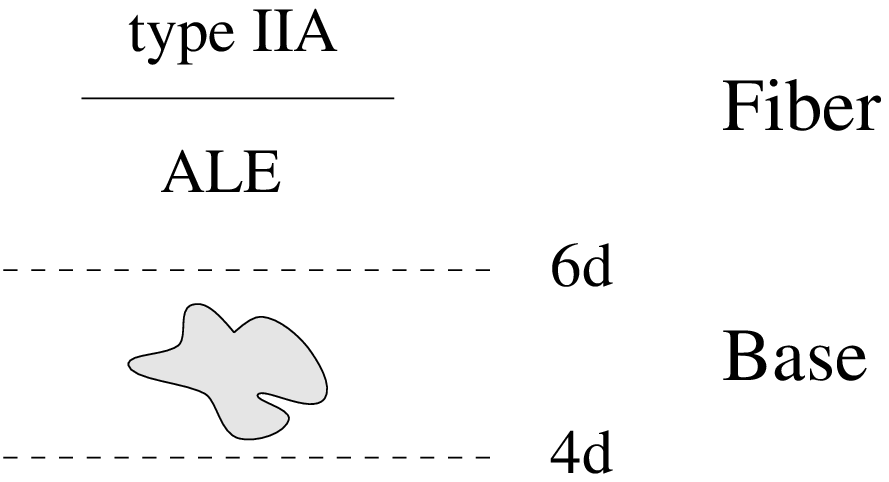}}
\leftskip 1pc\rightskip 1pc \vskip0.1cm
%\noindent{\ninepoint  \baselineskip=8pt {\bf Fig. 1:}}
\endinsert}\ni
\nin{\it $N=4$ Torus compactification}: 
The simplest example is a compactification where the base geometry is that
of a torus. In this case we get extra scalars from  decomposing the 
six-dimensional vector fields $A \to \omega^\prime \wedge \phi$, where
$\omega^\prime$ is a harmonic one-form on the torus. An important relation,
which exists independently of the special base geometry, is the
relation of the four-dimensional gauge coupling to the volume
of the base, as obtained from dimensional reduction:
\eqn\gci{
\frac{1}{g_6^2}\cdot Vol(Base)=\frac{1}{g_4^2}
}
This means in particular, that the Montonen-Olive duality
$g_4\to 1/g_4$ arises from the geometric T-duality $V\to 1/V$,
a classical string symmetry \refs{\duff,\witi,\vafai}. 
\mbr

\nin{\it $\IP^1$ as the base: $N=2$ in four dimensions}: To reduce
the supersymmetry in the compactification from six to four
dimensions (and therefore to end up with $N=2$ supersymmetry)
we have to choose a base geometry with non-trivial curvature.
The simplest case is to take again a two-sphere, or $\IP^1$. The
curvature of $\IP^1$ kills the extra scalars from the vector
multiplets; in particular there are no harmonic one-forms, $h^{1,0}(\IP^1)=0$. 
However to preserve the Calabi--Yau condition, the total 
geometry can no longer be of a simple product form; the
ALE space has to vary over different points on the base $\IP^1$.
This geometric structure is called a fibration, more precisely, in this
case we have a fibration of an ALE space over a base $\IP^1$.\mbr

\nin{\it Gauged coupling constants}: 
There is an interesting aspect of the geometric construction,
which will play a role later on in the case of four-dimensional
theories with vanishing beta-function. Consider the simplest 
configuration of a single $\IP_{Fiber}^1$ fibered over a base $\IP_{Base}^1$:
\vskip 0.5cm
{\baselineskip=12pt \sl
\goodbreak\midinsert
\centerline{\epsfxsize 2.5truein\epsfbox{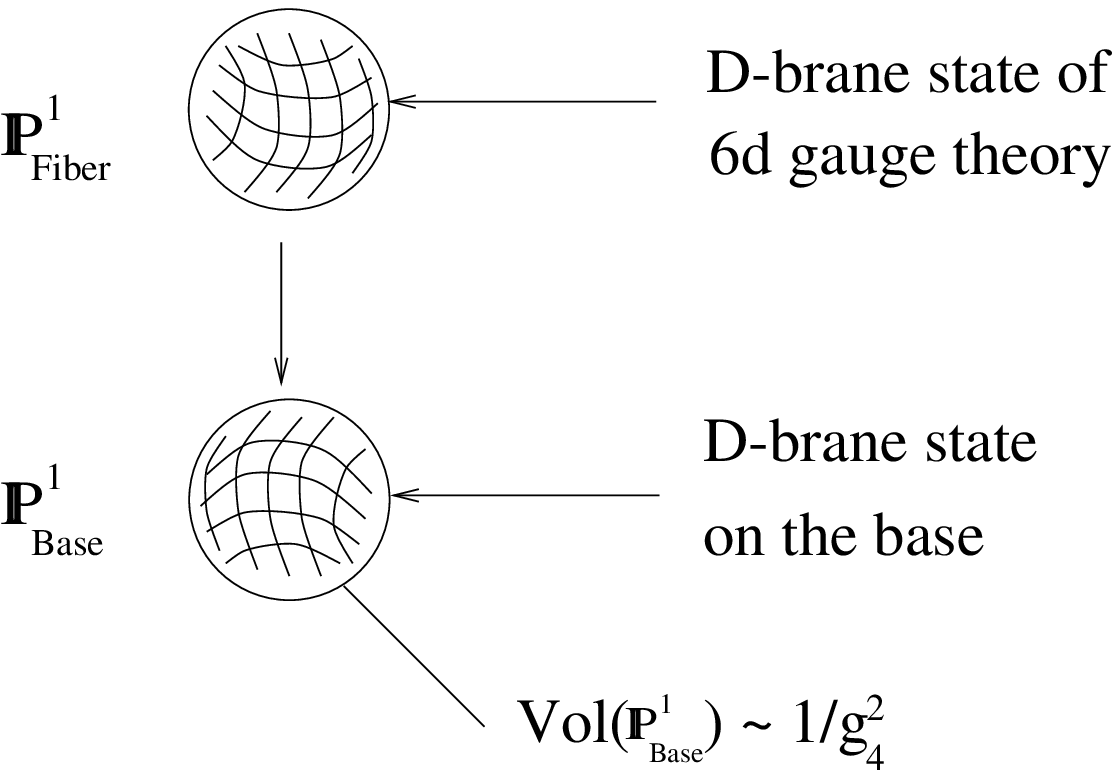}}
\leftskip 1pc\rightskip 1pc \vskip0.3cm
\noindent{\ninepoint  \baselineskip=8pt \centerline{{\bf Fig. 9:}
$SU(2)$ gauge theory from the base.
}}\endinsert}\ni
\nin As mentioned previously, we have still the relation 
$1/g_4^2 \sim Vol(\IP^1_{Base})$. However note that the coupling
$g_4$ appears as the scalar component of a full vector
multiplet. Moreover, from wrapping the D2-brane over the
base $\IP_{Base}^1$, we get charged $W^\pm$ vector multiplets as well.
These states are the light degrees of freedom of a different
$SU(2)_{Base}$ theory \ref\scs{A. Klemm and P. Mayr, \nup 469
(1996) 37;\br
S. Katz, D.R. Morrison and M. Plesser, \nup 477 (1996) 105.}, 
which however is restored  at infinite coupling  
of the six dimensional $SU(2)_{Fiber}$ from the fiber $\IP^1_{Fiber}$,
since its Coulomb parameter is related to the gauge coupling of the 
fiber theory by 
\eqn\couplii{
a(SU(2)_{Base}) \sim \frac{1}{g_4^2(SU(2)_{Fiber})}\ .
}
In general,
this gauge theory will decouple in the $m_{pl}\to \infty$ limit
which we take in going from string theory to the field theory
limit with gauge group $SU(2)_{Fiber}$. However this is not true
in the case of vanishing beta-function for the fiber gauge group.
This will lead to interesting new theories discussed later on.
\mbr

\nin
{\it Incorporation of matter}:
Let us next explain in more detail the appearance of 
matter representation in the four-dimensional $N=2$ theory.
Recall the logic of the geometric construction: the perturbative
gauge theory is defined by the charge lattice generated by the
roots (determining the gauge system) and now in addition 
weights for the matter representations. If this lattice is
realized geometrically as the lattice of homology two-cycles, 
the D-brane wrappings of the type IIA theory will generate
the appropriate physical states in the string compactification.
So to add matter, all we have to do is to add two-cycles which 
intersect in the appropriate way. The simplest example is 
again that of a $SU(2)$ theory, now with $N_f=1$ matter.
To obtain a matter multiplet, we simply add a new two-cycle,
a $\IP^1$ which intersects the first $\IP^1$ of the $SU(2)$
gauge theory.

The D-brane wrappings on the first $\IP^1$ still generate
the $W^\pm$ bosons of the $SU(2)$ gauge theory, while the
wrapping on the second, intersecting $\IP^1$ should correspond to the matter
multiplet. However such a 
configuration reminds very much of the geometry of a 
$SU(3)$  theory as in fig.2. In fact this is true
up to a small subtlety: the new $\IP^1$ which provides the matter is
${\it localized}$ on the base $\IP^1$. In more detail this
means that over the generic point $z$ of the base $\IP^1$, there
is only a single two-cycle in the fiber which supports the 
gauge bosons, while for a special point on the base $\IP^1$,
say $z=0$,
the fiber contains an extra two-cycle class that supports the matter.
This geometry is shown in fig.4.

The fact that this geometry is similar to the geometry of 
a $SU(3)$ theory is related to the fact that the matter
content of geometrically constructed $N=2$ theories can be understood 
in terms of adjoint breaking \kvii. Consider the breaking 
of the $N=2$ gauge theory in six dimensions by vev's of the adjoint 
scalar fields. The idea is to consider fibrations, where the scalar 
field of a $U(1)$ subgroup of the gauge group $G\ovset H\times U(1)$
is identified with the fibration parameter $z$. 

The surviving gauge
group in the lower-dimensional theory is $H$. Over a general point
on the base, the $G$ singularity of the fiber is resolved to 
an $H$ singularity and the two-cycle classes of the latter support
the vector bosons corresponding to the roots of $H$. However at the
special point $z=0$, the singularity is still of type $G$ and the
extra, localized two-cycles give rise to additional states in a
representation $R^\prime$ of $H$. Here $R^\prime$ denotes the 
representation obtained by the decomposition of $adj(G)$ according
to the breaking $G\ovset H\times U(1)$. E.g., in the above example 
we have
\eqn\xxx{
SU(3)\supset SU(2)\times U(1):\ \ul{8}\to\ul{3} + 2\cdot\ul{2}+\ul{1} \ ,
}
that is $R^\prime=\ul{2}$, in agreement with the appearance of 
a fundamental hyper multiplet.

The Lorentz quantum numbers of the states wrapped on the generic or
special $\IP^1$'s follow from the quantization of the collective coordinates
corresponding to the moduli space of the two-cycles \refs{\witm,\kvii}. 
Note that the moduli space of a generic $\IP^1$ is a $\IP^1$ 
(the base) and that of a special $\IP^1$ is a single point.
A heuristic explanation follows from a brane picture using open strings,
either in F-theory \kvii\ or from the T-dual configuration \seni\
of flat branes. E.g., in the latter case it is well-known,
that parallel branes lead to enhanced non-abelian gauge
symmetries, while intersecting branes generate matter \vafaii.
Note that the gauge bosons arise from open strings which 
can move freely along all directions of the parallel branes
whereas the open string between intersecting branes is localized
in the directions which are not common to both branes. 
In this case the determination of the Lorentz quantum numbers,
arising from a quantization of fermionic zero modes, is
identical to a simple orbifold calculation \vafaii.
\mbr

\nin{\it Product gauge groups with bi-fundamental matter}:
As is clear from the relation to adjoint breaking, the 
above example is actually a simple subcase of a more general
class of geometries which give rise to product gauge groups
with matter representations determined by adjoint breaking \kvii.
E.g. instead of $SU(N)\supset SU(N-1)\times U(1)$, we can consider
a breaking $SU(N)\supset SU(K)\times SU(N-K)\times U(1)$.
In more general terms we consider collisions of any ADE singularities
on the base manifold. Specifically, consider the case
where we have a curve of $A_N$ singularities (that is a
base $\IP^1$ above which the fiber has a singularity of
type $A_N$) intersecting with a curve of $A_M$ singularities.
Note that the base consists now of {\it two} $\IP^1$ factors,
one for the $A_N$ singularity and one for the $A_M$ singularity.
These two $\IP^1$'s intersect at a point. The "Dynkin diagram"
of the base geometry is therefore that of an $A_2$ singularity.
\vskip 0.5cm
{\baselineskip=12pt \sl
\goodbreak\midinsert
\centerline{\epsfxsize 2.0truein\epsfbox{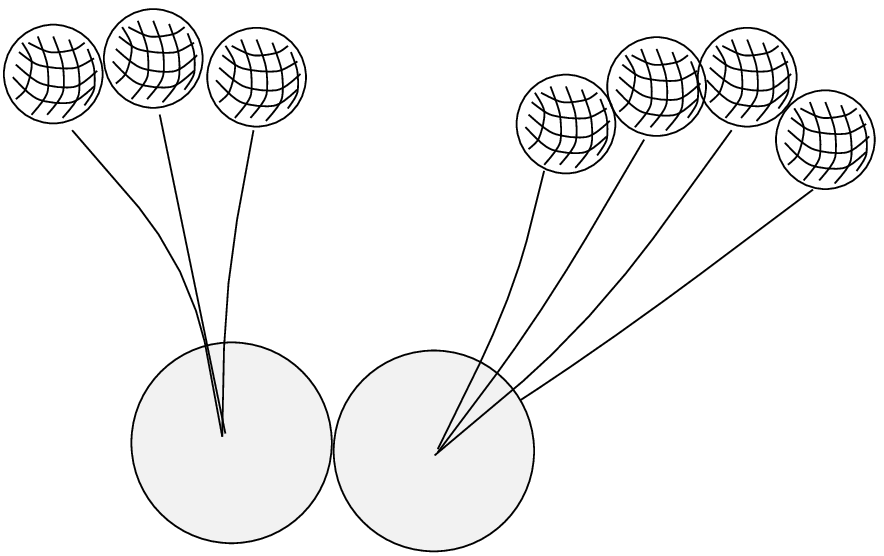}}
\leftskip 1pc\rightskip 1pc \vskip0.3cm
\noindent{\ninepoint  \baselineskip=8pt \centerline{{\bf Fig. 10:}
Intersection of two $A$ type singularities in the fiber with an $A_2$ base 
geometry.
}}\endinsert}\ni
\nin
A general mathematical result assures that at the intersection
point, the fiber singularity is of type $A_{N+M+1}$. 
In other words, at the intersection point, there is an extra
two-cycle class corresponding to the $+1$. This is the localized
$\IP^1$ which carries a matter multiplet in the $(\ul{N+1},\ul{M+1})_{q(U(1))}$
representation of the $SU(N+1)\times SU(M+1) \times U(1)$ gauge group
(fig.11).
\vskip 0.5cm
{\baselineskip=12pt \sl
\goodbreak\midinsert
\centerline{\epsfxsize 3.8truein\epsfbox{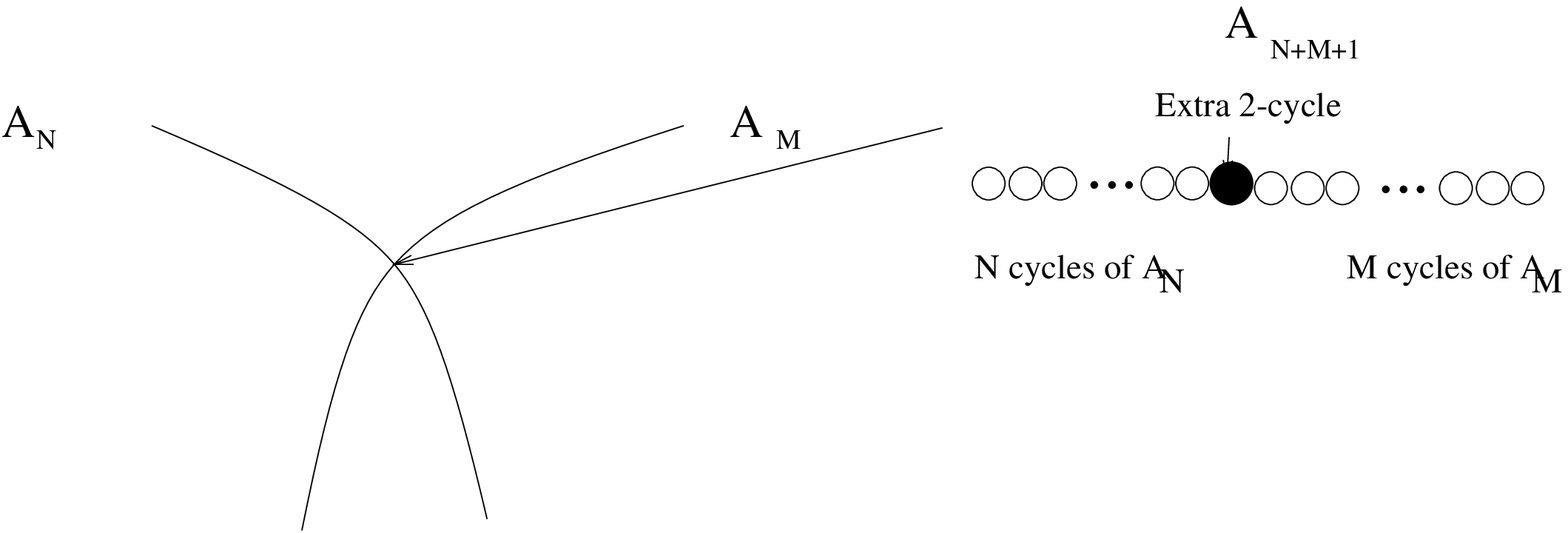}}
\leftskip 1pc\rightskip 1pc \vskip0.3cm
\noindent{\ninepoint  \baselineskip=8pt \centerline{{\bf Fig. 11:}
Localized bi-fundamental from enhancement of the singularity.
}}\endinsert}\ni
\nin
Note that the bare mass of the matter multiplet
corresponds
to the Coulomb parameter of the extra $U(1)$ factor. Similarly as
in the case of the gauge coupling, this bare mass is part of a full
vector multiplet and is therefore gauged. 
\mbr

\nin{\it Degenerate factors: fundamental matter}:
It is now easy to construct geometries leading to $SU(N)$ 
factors with $M$ fundamental matter multiplets: recall that
the gauge coupling of the $SU(M+1)$ theory in the previous 
paragraph is given by the volume of the corresponding 
base $\IP^1$, $1/g_{SU(M+1)}^2\sim Vol(\IP^1_{Base})$.
We can decouple the $SU(M+1)$ factor by making the second base
$\IP^1$ very large and therefore $g_{SU(M+1)}\to 0$. In this
limit the vector multiplets decouple, but the matter multiplets do not.
What remains geometrically is a single compact two-cycle in the
base with one extra special point, the former intersection point, 
above which there are $M+1$ extra two-cycle classes carrying 
the $M+1$ fundamental matter multiplets of the $SU(N+1)$ gauge
theory.
\mbr

\nin{\it General base geometries}:
As we mentioned already, the ADE singularities are the only possible
fiber geometries as a consequence of the classification of 
K3 singularities. This does not say anything about the
base geometry, however. In the absence of adjoint matter, the 
homology of two-cycles of the base geometry is again generated
by a collection of intersecting $\IP^1$'s which can be again
characterized by their intersections summarized in 
"generalized Dynkin diagrams"
as in fig.5. If we add the information about the 
fiber, we consider collisions of ADE fiber singularities,
described by these intersections of the base $\IP^1$'s
as in fig.11. At the intersection points
we should get matter representations charged under the 
gauge group factors corresponding to the intersecting fiber singularities.
\vskip 0.5cm
{\baselineskip=12pt \sl
\goodbreak\midinsert
\centerline{\epsfxsize 6truein\epsfbox{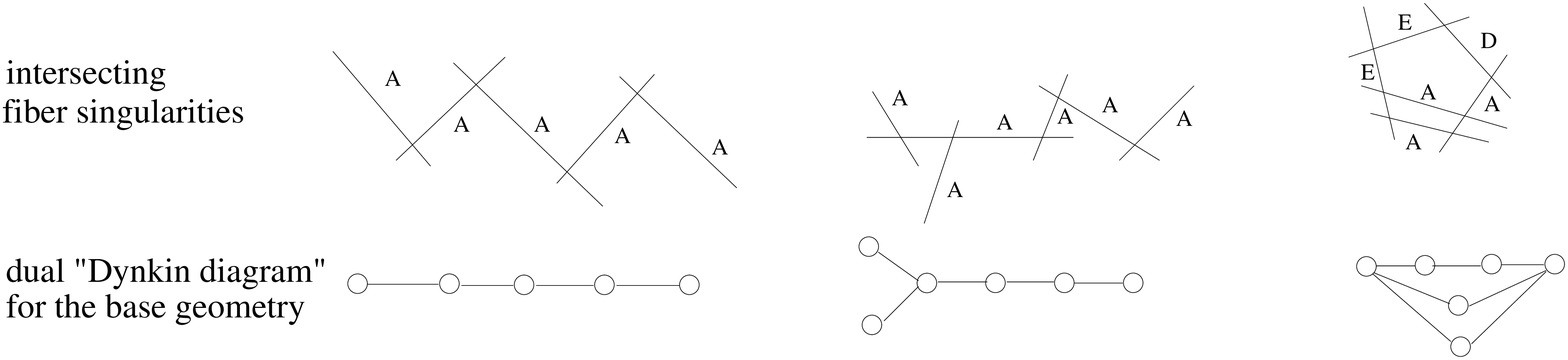}}
\leftskip 1pc\rightskip 1pc \vskip0.3cm
\noindent{\ninepoint  \baselineskip=8pt \centerline{{\bf Fig. 12:}
Intersections of fiber singularities and dual "Dynkin diagrams".
}}\endinsert}\ni
\nin
However not all of these intersections make sense in terms
of four-dimensional field theories, as is clear from the
fact that not all combinations of group factors meeting at the
intersection points can be obtained from adjoint breaking of a 
larger group. Only the two left diagrams in fig.12 
make sense in this class. Even if one gives up the constraint to obtain
a conventional field theory, not all possible collisions
will lead to theories which allow for a decoupling of 
gravity\foot{For criteria about the existence of such a limit, see
\lmw.}.
The classification of geometries corresponding to these two 
classes of $N=2$ theories is still an open question.
However there is a nice result for the subclass of field theories
with gauge group an arbitrary product of $SU(n)$ factors and
asymptotic free
bi-fundamental and fundamental matter representations (corresponding
to geometries with only $A$ type singularities in the fiber): the
only possible base geometries are configurations of 2-cycles
which intersect according to affine ADE Dynkin diagrams \kmv.

\subsec{Instanton corrections}
Given an appropriate geometry with a homology lattice
of two-cycles, the fundamental type IIA string together with the 
D brane states will give rise to
the physical states of an $N=2$ theory. We are 
interested now in getting the exact instanton corrected
effective action of this theory.

The determination of (an infinite number of) instanton
corrections to field theory is a very hard question.
As mentioned already, the case of $N=2$ supersymmetry can
be often solved starting from the knowledge of the perturbative
theory and requiring consistency of the solution
with holomorphicity and positivity of the gauge coupling.
String theory provides an alternative framework, which is
probably the most systematic and most physical one: the use of
mirror symmetry, a symmetry of classical string theory. 

That classical type II string theory can provide the exact solution
is due to the following two facts:

\item{o} there are no space-time
instanton corrections to the vector multiplet moduli
space. 
\item{o} there {\it are} worldsheet instanton corrections which are
non-perturbative from the string sigma model point of view.
However these instantons can be determined by mirror symmetry.
\mbr

\nin{\it Space-time instanton corrections}: 
Let us first recall briefly, why we do not have to bother
about space-time instanton corrections (from the point of
string perturbation theory). What we are interested in 
is the exact moduli space of the scalar
components $a_i$ of the neutral vector multiplets, 
which parametrize the flat directions on the Coulomb
branch of the $N=2$ theory. To obtain 
the exact gauge coupling $\teff(a_i)$ of the theory we do not have to
care about scalars in hyper multiplets, since there are
no neutral couplings between hyper multiplets and vector multiplets
in the $N=2$ supersymmetric theory \dec. 

This decoupling between hyper and vector multiplets is precisely the
reason for the absence of space-time instanton corrections
to the vector multiplet moduli: for the type II string compactifications
on Calabi--Yau manifolds which we consider, the 
string coupling constant $g_{string}$
appears as a real scalar in the dilaton hyper multiplet.
Therefore the gauge coupling constant on the Coulomb branch
does not depend on the string
coupling constant at all. Note that the
vector multiplets arise from the RR sector of the theory (the
anti-symmetric one and three-form potentials in ten dimensions). 
There are no fundamental states of type II string theory
which are charged under the RR gauge fields. The absence
of fundamental charged states, which have masses and couplings
that depend on $g_{string}$, may serve as a heuristic, physical
reasoning for the absence of space-time instantons.

The fact that there are no space-time instanton corrections
to the vector multiplet moduli space in the string theory 
makes the determination of the exact result of course much easier:
all we have to calculate is the tree-level string theory answer.
\mbr

\nin{\it Geometrical instantons}:
Given the generically infinite series of instanton contributions
to $N=2$ SYM field theory however, it is clear that there
has to be some source for these non-trivial corrections
in string theory - if its answer succeeds
to reproduce field theory in the appropriate point-particle
limit. These corrections arise from worldsheet instanton
corrections, that is corrections which are non-perturbative
from the string sigma model point of view. Again they can
be understood as wrappings of supersymmetric two-cycles of
the compactification geometry; this time however we consider
euclidean wrappings of
the $1+1$ dimensional fundamental string worldsheet rather
than wrappings of D2-branes. 

Consider again the simplest situation, where we have a two-sphere
$\IP_{Fiber}^1$ fibered over another two-sphere $\IP^1_{Base}$.
Moreover we consider a euclidean string worldsheet, which is
wrapped $k$ times on the fiber $\IP^1$ and $m$ times on the
base $\IP^1$.
\vskip 0.5cm
{\baselineskip=12pt \sl
\goodbreak\midinsert
\centerline{\epsfxsize 3.2truein\epsfbox{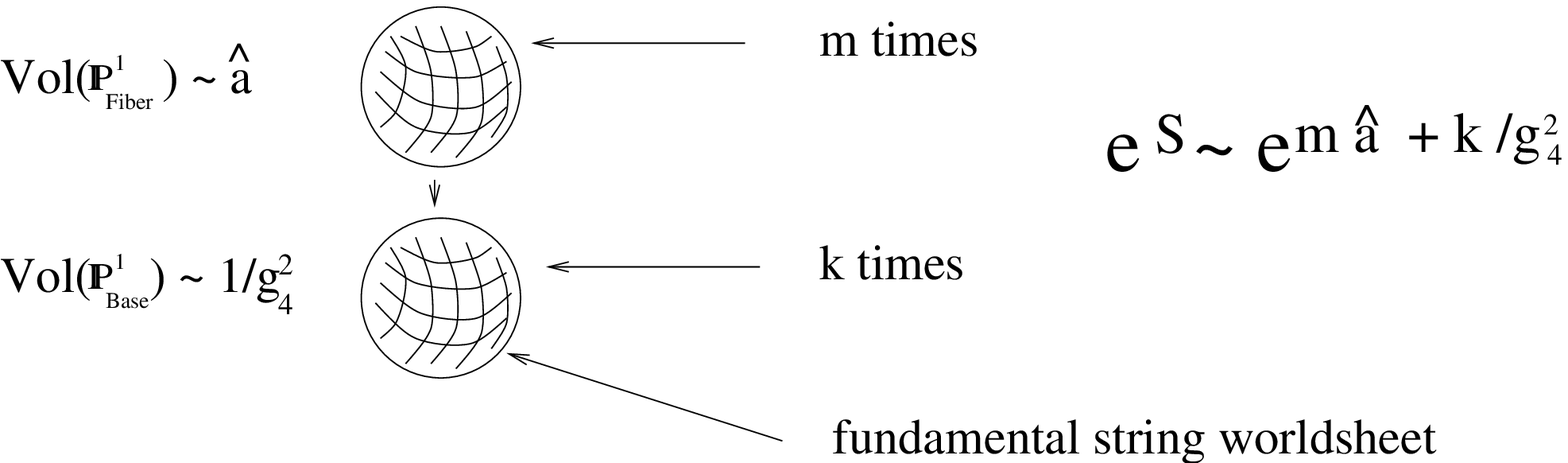}}
\leftskip 1pc\rightskip 1pc \vskip0.3cm
\noindent{\ninepoint  \baselineskip=8pt \centerline{{\bf Fig. 13:}
Worldsheet instanton.
}}\endinsert}\ni
If $B$ denotes the class of the base $\IP^1$ 
in $H_2(X_3)$ and $F$ that of the fiber $\IP^1$, then the
class of $C_2$ is $C_2=k\cdot B + m \cdot F$ and the 
instanton action of this wrapping is 
\eqn\wsw{
S\sim Vol(C_2)=k\cdot Vol(\IP^1_{Base}) + m \cdot Vol(\IP^1_{Fiber})=
{\rm const.} \frac{k}{g^2}+m\hat{a} \ ,
}
where we have used eq.\gci\ which identifies $Vol(\IP^1_{Base})$
with the gauge coupling and moreover $\hat{a}$ denotes the 
scalar field which measures the volume of the fiber $\IP^1$
and is related to the Coulomb scalar of the field theory
vector multiplet by a holomorphic redefinition. Thus 
a worldsheet instanton wrapped $k$ times on the base has
an action with the same dependence on the gauge coupling as a
$k$ space-time instanton from 
the point of the gauge theory.
\mbr

Now the direct calculation of the contributions of infinitely many
different worldsheet wrappings would be similarly hopeless
as the calculation of the space-time instantons directly in field theory.
This is the point where mirror symmetry comes to help \cani. It maps 
the worldsheet instanton corrected 
type IIA theory, which we used to generate the perturbative
spectrum via D2-brane wrappings, to a type IIB compactification
on a different manifold. In the latter theory, the worldsheet
instantons do not correct the vector moduli space. To explain
this step, let us recall some facts about the moduli space
of Calabi--Yau compactifications.
\mbr

\nin{\it Calabi--Yau moduli spaces}: A Calabi--Yau three-fold $X_3$
is a three complex dimensional K\"ahler manifold with vanishing first 
Chern class. The latter condition implies the existence of a
covariantly constant spinor field which gives rise to an $N=2$
supersymmetry in the type II compactification to four dimensions.
The generic holonomy group is $SU(3)$; if the
holonomy group is further reduced, as in the case of $K3\times T^2$ or 
$T^6$, there are more than one covariantly constant spinors and
the four-dimensional supersymmetry algebra is extended to
$N=4$ and $N=8$, respectively \ref\canlec{For an excellent lecture see
P. Candelas, {\it Lectures on complex manifolds}, proceedings, Trieste
1987.}. 

\ni
There are two types of parameters which describe the geometry
of $X_3$: \vskip0.2cm
\item{} {\bf K\"ahler moduli} (KM) $t_a$ are defined in terms of volumes  of
holomorphic two-cycles $C_2^a\in H_2(X_3)$. If $J$ is the K\"ahler form
on $X_3$, then the value of $t_a$ is given by $t_a=\int_{C_2^a}J$.
These parameters can be understood as measuring the sizes of $X_3$
and analytic submanifolds in $X_3$.

\item{}{\bf Complex structure moduli} (CSM) $z_i$ are defined in terms
of the volume of three-cycles $C_3^i\in H_3(X_3)$. The volume form is
given by the unique holomorphic tree-form $\Omega$; a convenient
parameterization of the complex structure is in terms of the 
{\it period integrals} of $\Omega$
\eqn\tfp{
z_i=\int_{C_3^i}\Omega \ .
}

In type II string compactifications, these parameters appear
as the scalar components of vector or hyper multiplets in the 
following way:

$$
\vbox{\offinterlineskip\tabskip=0pt\halign{\strut
\hfil ~#~ \hfil &\vrule#
&\hfil ~#~ \hfil &\vrule#
&\hfil ~#~ \hfil \cr
&& \ \ {\it type IIA} && \ \ {\it type IIB} \cr
\noalign{\hrule}
\ \ {\it vector multiplets}
&& \ \ K\"ahler structure && \ \ complex structure \cr
\ \ {\it hyper multiplets} && \ \ complex structure, 
$g_{string}$ && \ \ K\"ahler structure,
$g_{string}$\cr
}}
$$

From the above table it is now obvious, that there are no 
worldsheet instanton corrections to the vector multiplet in
the type IIB theory for the same reason that there are
also no space-time instantons\foot{The string coupling constant
is again part of a hyper multiplet in type IIB; thus there are
no space-time instantons corrections to the type IIB vector
multiplet moduli either.}: the worldsheet instanton action
depends on the volume of the relevant two-cycle $S\sim Vol(C_2)$
as in \wsw\
which  corresponds to a scalar field in the K\"ahler moduli
of $X_3^*$. However the supersymmetric multiplet of the 
K\"ahler moduli is the hyper multiplet in the type IIB theory.
Therefore a dependence of the vector multiplet moduli space
on the worldsheet instantons would contradict the decoupling
of hyper and vector multiplets. 
\mbr

\nin
{\it Type IIB (D brane-) geometry}: Mirror symmetry maps
the type IIA compactification on a Calabi--Yau threefold
$X_3$ to a type IIB compactification on the mirror manifold
$X_3^*$ and the moduli space $\cx M(X_3)$ to the moduli
space $\cx M(X_3^*)$ with K\"ahler and complex structure 
moduli exchanged. Moreover it maps the D brane
states of the type IIA theory to D brane states of a type IIB
theory.

The even-dimensional D branes of type IIA theory 
give rise to point particles when wrapped
on supersymmetric d=0, 2, 4, 6 cycles in $X_3$. The states
which carry perturbative charges with respect to the
gauge fields $A^a$ obtained from the decomposition $A^{(3)}=
A^a \wedge \omega^a$ arise from the D2-branes wrapped 
on two-cycles $C_2^a$. This is the reason why we could concentrate
on the geometry of two-cycles in the previous discussions.
The magnetic-electric dual states arise from the dual homology\
cycles, that is D4 branes wrapped on 4-cycles\foot{In addition
there is the universal vector multiplet which exists already
in ten dimensions, corresponding to the universal 0/6 cycle
on $X_3$.}.

Mirror symmetry maps all these even dimensional branes of
type IIA to D3 branes of type IIB wrapped on three-cycles. 
A hint that the type IIB description is appropriate
for the analysis of non-perturbative effects comes
from the fact that now electric and magnetic states
are described by the same object, a wrapped D3 brane. 
In fact it is also easy to see from the D brane point of 
view, why the type IIB theory is free of worldsheet
instanton corrections and the classical geometrical 
answer will agree with the exact result: the
scalars of the vector multiplets parameterize 
now volumes of three-cycles rather than volumes of two-cycles.
An instanton correction from wrapping an extended object 
will be proportional to the volume of these three-cycles.
However there are no appropriate two dimensional branes
in the type IIB theory which could be wrapped on the 
three-cycles (the fundamental string worldsheet wrappings 
give again rise to corrections of the K\"ahler moduli 
space, which is now parameterized by the hyper multiplets,
however).

Now recall that the vector multiplets correspond to the 
CSM in type IIB which in turn are parameterized by the period integrals
of $\Omega$ \tfp. Since there are neither perturbative nor
instanton corrections at all,
{\it the classical period integrals of $X_3^*$ describe the exact
vector multiplet moduli space of the $N=2$ theory in four dimensions}.
This is of course very similar to the situation observed in \sw,
were the exact solution of $N=2$ $SU(2)$ theory has been found
to be given in terms of period integrals on a torus \swsii. We will
explore this relation in more detail below with the result that

\vskip 0.1cm {\it
The general solution to a $N=2$ SYM theory is given in terms
of period integrals $\int_{C^i_3} \Omega$ on a local 
Calabi--Yau geometry  $S\subset X_3^*$.}
\vskip 0.1cm

\nin
Here $\Omega$ is again the unique holomorphic three-form of $X_3^*$.
In special cases, such as the $SU(2)$
example, it can be  natural to do part of the three dimensional integration
and to obtain in this way a description in terms of period integrals
on a one-dimensional geometry, the Seiberg--Witten torus $\Sigma$.

\subsec{Four-dimensional field theories from the six-dimensional string}
We want now to take more serious the point of view to consider
the four-dimensional theory obtained from type II on a Calabi--Yau
three-fold as a six-dimensional compactification followed by a 
further compactification to four dimensions. This viewpoint
is particularly interesting from the point of the type IIB theory.

Before we explain in more detail the geometry of the type IIB
compactification on $X^*_3$, which is mirror to the type IIA 
compactification on $X_3$ discussed above in detail, 
let us see what one expects 
intuitively. In the type IIB theory we start with the self-dual
string in six dimensions, which is obtained from wrapping the
D3 brane on one of the two-cycles $C_2^a$ of a K3 or a non-compact 
Calabi--Yau two-fold with $SU(2)$ holonomy. In the four-dimensional
$N=4$ compactification on $K3\times \IT^2$, the winding states
of these strings on the torus give rise to point like degrees
of freedom. More precisely, if $\alpha_i,\ i=1,2$ is a standard
basis of 1-cycles on the extra $\IT^2$, we get electric (magnetic)
states from wrapping the string around $\alpha_1$ ($\alpha_2$). Moreover,
the T-duality transformation $Vol(\IT^2)\to 1/Vol(\IT^2)$ of the
torus interchanges $\alpha_1$ and $\alpha_2$, This is of course
consistent with the fact that this T-duality represents 
an electric-magnetic duality transformation in the $N=4$  field theory.
Note that the type IIB D3 brane is now wrapped 
on a three-cycle $C^a_2\times \alpha_i$

In the $N=2$ case the situation is very similar. Similarly as
before, we can consider a D3 brane wrapping on three-cycles $C_3^i$,
which are roughly speaking composed of a D3 brane wrapped on a 
two-cycle in the ALE fiber
and a winding of the resulting string on the base (which 
will turn out to be the Seiberg--Witten geometry $\Sigma$ for $G=A_n$). 
Again these winding 
states give rise to the charged states of the four-dimensional theory.
Note that  the gauge bosons and monopoles are treated on equal footing
similarly as in the $N=4$ case.
\mbr

\nin{\it Mirror geometry of ALE fibrations}: To get a more complete 
picture, let us describe qualitatively the action of mirror symmetry
on the geometry of the type II compactifications.

In the type IIA theory, we start with a six-dimensional compactification
on the ALE space of ADE type.
In the compactification to four-dimensions on a further $\IP^1$, the
total space can not be a simple product in order 
to satisfy the Calabi--Yau condition. 
Rather the complex structure of the ALE fiber varies over the 
points on the base $\IP^1$. However the volumes of supersymmetric
two-cycles, which are holomorphic in the complex structure of the
three-fold $X_3$, do {\it not} vary. This geometric structure is of the form
of a holomorphic fibration (the ALE fiber corresponds to a certain divisor 
class of $X_3$). 
Each of the holomorphic two-cycles which stems
from the fiber, supports a D2-brane state of six dimensional origin
in the four-dimensional $N=2$ theory. 

Note that we could have been more general in choosing the base geometry
and in particular we can choose the two-cycle homology from the {\it base} to
generate the intersection lattice of an ADE singularity
corresponding to any (affine) ADE Dynkin
diagram\foot{The example of a single base $\IP^1$ corresponds to the
$A_1$ case.}. 

This will be important due to the following fact: as is well known,
the K3 manifold is self-mirrored. In fact, since we are interested
in singularities of K3, we can use a local version of this statement:
the mirror of a deformation of an ADE singularity of type $G$ is again 
a deformation of an ADE singularity of type $G$. More precisely, 
a K\"ahler deformation is mapped to a deformation in complex structure.

Since our local fiber geometry is of the form of 
an ALE space with ADE singularity and the base geometry will
correspond to the resolution of an ADE singularity as well,
mirror symmetry acts somehow trivially on both the 
fiber and the base. Only the fibration, that is
the variation of the fiber over the base, {\it is}
affected non-trivially by mirror symmetry.

In fact the ``fibration'' structure of the type IIB geometry $X_3^*$
is rather different from that of $X_3$ and in general is not
of the type of a holomorphic fibration. In the type IIB geometry,
though there is still an ALE space over each point on the
base geometry, the size of the relevant two-cycles varies with
the point on the base. We will continue to use the notation of an
ALE fiber and the base for the type IIB geometry though it 
is not a fibration in the usual sense. 
In particular there are special points
on the base where the volume of two-cycles $C_2^a$ in
the fiber vanishes. 
So although the naive mirror of the base geometry, which is
of the form of a resolved  ADE singularity, is again a resolution of 
an 
ADE singularity
of the same type, there are now extra points $p_a$ on this base geometry
above which a special two-cycle $C^a_2$ vanishes in the ALE fiber space.

This is important due to the fact that vanishing cycles in 
the ALE fiber will be associated with non-trivial {\it monodromies}.
Let us explain this effect by a simple example. Given a complex torus $\Sigma$,
one can define a basis of one-cycles $\alpha_i,\ i=1,2$. The 
complex structure $\tau$ of $\Sigma$ can then be expressed in terms of the 
period integrals $\Pi_i=\int_{\alpha_i}\omega$ as $\tau=\Pi_1/\Pi_2$.
Here $\omega$ is the holomorphic one-form on $\Sigma$. 
However if we move to a point
in the moduli space where one of the two-cycles, say $\alpha_1$, vanishes,
the definition of the basis becomes ambiguous since we can add multiples
of $\alpha_1$ to $\alpha_2$ and switch the sign of $\alpha_1$ 
without changing the complex structure $\tau$. In general,
if we move in the moduli space, that is the $\tau$ plane divided by discrete
identifications, around a closed path that includes
a point where one of the cycles vanishes, the basis will be redefined
by additions of the vanishing cycle and possibly minus signs. 
This effect is called monodromy. 
%
%\begin{figure}[ht]
%\vspace{-0.8cm}
%\hbox to\hsize{\hss
%\epsfysize=1.5cm
%\epsffile{fig18.eps}\hss}
%\caption{Monodromy for the torus}
%\end{figure}
%\vspace{0cm}
%
%\nin

Precisely this situation appears in the type IIB geometry: the 
two-cycles of the ALE space vary with the position $z$ on the base,
which therefore plays the role of the modulus in the above torus
example. Moving around a closed path on the base which encircles a
point where a two-cycle vanishes in the ALE space, the basis of the
homology lattice of two-cycles of the ALE space is affected by a
redefinition. Though the totality of two-cycles and their intersection
properties do not change of course, individual cycles
may be exchanged and redefined. The intersection lattices
which we consider correspond to root lattices described by
(affine) Dynkin diagrams of ADE. These lattices are invariant under the
appropriate Weyl group. Since the monodromy has to leave invariant
the total lattice, the monodromy transformations act as Weyl transformations
on the homology of two-cycles of the ALE space. 

To recap, though the mirror transformation acts somehow trivially
on the fiber and the base as a consequence of the self-mirror 
property of the ADE singularity, the fibration structure changes.
In particular there are now points on the base where some two-cycle
volumes in the ALE fiber vanish. These points are associated with monodromies
which take values in the Weyl group of the ADE fiber singularity. 

Note that the base $B$ is now described by a collection of intersecting
$\IP^1$'s with extra points around which there are monodromies. 
Alternatively, we could consider a multiple cover $\hat{B}$ of $B$ such that
a closed path on $\hat{B}$ has trivial monodromy. This is the definition
of a Riemann surface. In fact, in the case of $A_n$ these Riemann surfaces
are precisely of the form which has been obtained in field theory from
consistency reasonings \refs{\sw,\suo}. 
\vskip 0.5cm
{\baselineskip=12pt \sl
\goodbreak\midinsert
\centerline{\epsfxsize 4.5truein\epsfbox{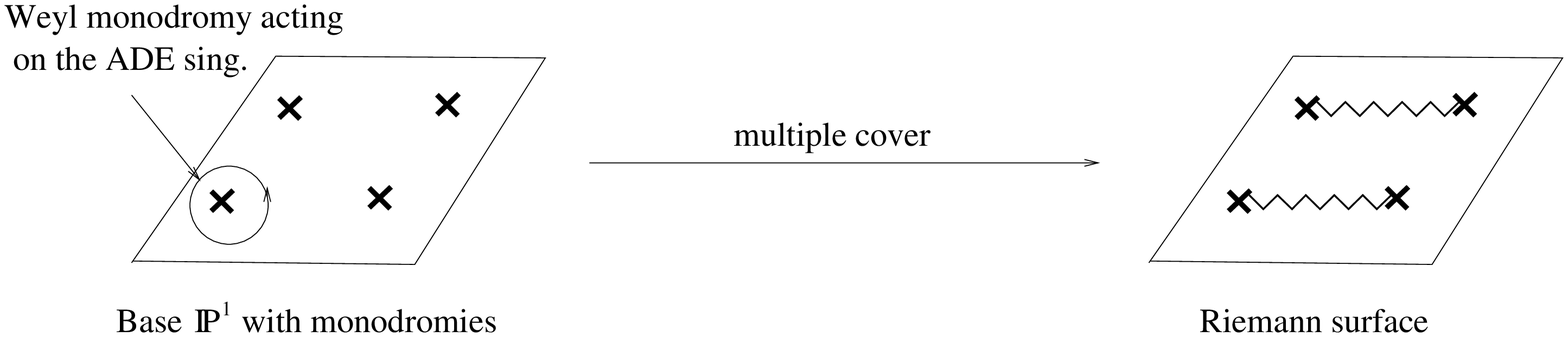}}
\leftskip 1pc\rightskip 1pc \vskip0.3cm
\noindent{\ninepoint  \baselineskip=8pt \centerline{{\bf Fig. 14:}
Riemann surface $\Sigma$ as a multiple covering of the $z$ plane.
}}\endinsert}
\vskip-0.2cm
\subsec{Decoupling limit and the dimension of the Seiberg--Witten 
geometry}
The exact vector moduli space obtained from the type IIB theory on 
Calabi--Yau three-fold $X_3$ contains the information about all
gravity and string effects in the $N=2$ theory in four dimensions.
Let us discuss in more detail the geometrical limit which decouples
these effects.

As an example consider the asymptotic free $SU(2)$ SYM theory.
Physicswise, we want to decouple the massive string and gravity states
by sending the string scale $m_{string}$ to infinity, however at fixed
strong coupling scale 
$\Lambda$ and fixed vector boson mass, $m_{W_\pm}$.
\def\ap{\alpha^\prime}
\eqn\ppl{
\ap \sim m_{string}^{-2}\to 0,\qquad
\Lambda\sim {\ap}^{-1/2}e^{-\frac{1}{bg^2}} \sim {\rm const.},\qquad
m_{W_\pm}/m_{string}\sim {\ap}^{1/2}\to 0 \ .
}
From eqs.\ppl,\gci\ we learn that the appropriate limit for
the volumes of the fiber and the base is
\eqn\vols{
Vol(\IP^1_{base})\sim -\frac{b}{2} \ln \ap \to \infty ,\qquad
Vol(\IP^1_{fiber})\sim {\ap}^{1/2} \to 0 \ .
}
Therefore we have to consider a geometrical limit of {\it large base 
and small fiber}. Note that taking a large base corresponds to a very small 
gauge coupling constant $g$ at the string scale $m_{string}$. This is
necessary to keep the strong coupling scale $\Lambda$ fixed, taking into
account the running of the coupling constant from $m_{string}$ to the
low energy scale of the field theory. Note also that the large base limit at the 
same time freezes the dynamics of the gauge theory from the base described
below fig.9; in particular the mass of the D-brane
wrappings on the base diverge in this limit.

A very special situation arises for $N=2$ theories with vanishing 
beta-function coefficient $b=0$. In this case there is no need
to take a large base limit to keep the gauge theory coupling finite
in the limit $m_{string}\to\infty$. If we keep the base volume of the order
of the fiber volume, the D-brane states wrapped on the base
become equally relevant as the states from wrapping the fiber two-cycle.
This gives rise to a interesting kind of interacting conformal
$N=2$ theories in four dimensions \kmv.
\mbr

\nin{\it The dimension of the mirror geometry $S$}: Taking the
geometrical limit described above, we obtain the effective 
action of a gravity free $N=2$ gauge theory in terms of 
periods of the holomorphic three-form on a local Calabi--Yau three-fold
$S\subset X_3$. However the original solution of Seiberg and Witten is
presented in terms of period integrals of a one-form on a 
Riemann surface. What is the explanation for this reduction of 
dimension of the mirror geometry ?

The answer is that in general the "Seiberg--Witten geometry" $S$ {\it is}
a Calabi--Yau three-fold. Only for sufficiently simple gauge groups 
such as $SU(n)$ is it natural to integrate out two of the dimensions 
to obtain a representation in terms of period integrals on 
Riemann surfaces $\Sigma$. Note that the reduction of the dimension is
not a consequence of the $m_{string}\to \infty$ limit. Rather we
get still local three-fold singularities of vanishing three-cycles
as the mirror geometry. In the general case, such as for a $E_8$ gauge
group or product gauge groups, this is the most natural answer.
\vskip 0.5cm
{\baselineskip=12pt \sl
\goodbreak\midinsert
\centerline{\epsfxsize 4.5truein\epsfbox{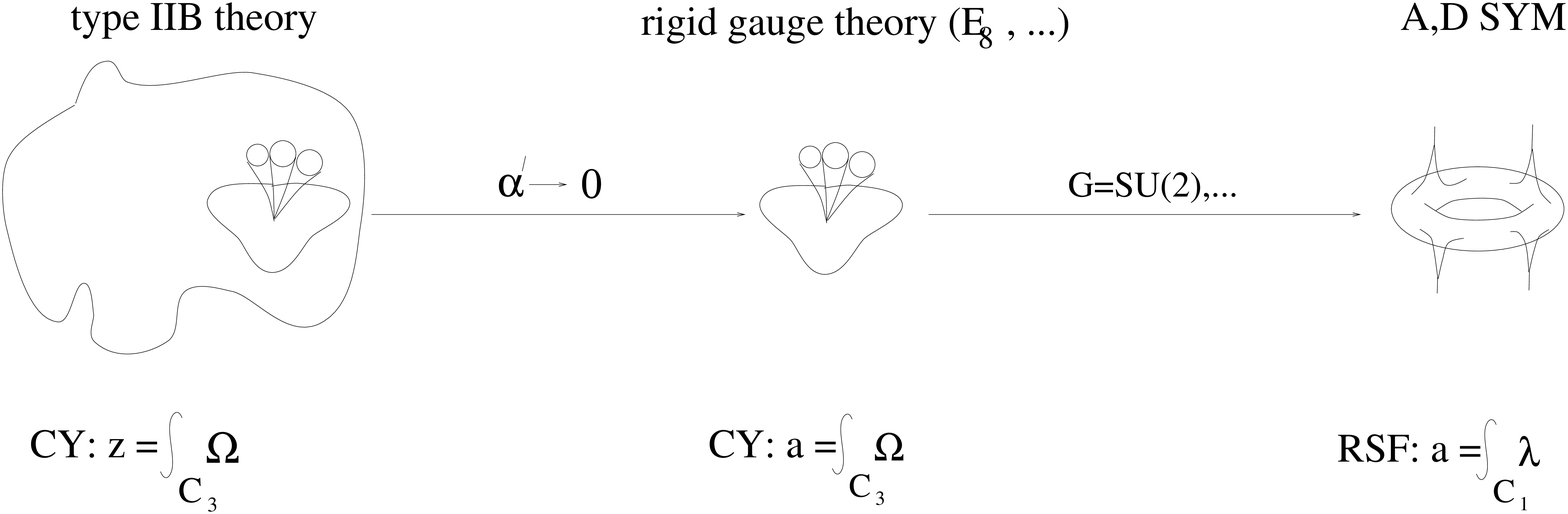}}
\leftskip 1pc\rightskip 1pc \vskip0.3cm
\noindent{\ninepoint  \baselineskip=8pt \centerline{{\bf Fig. 15:}
Geometrical limits of the mirror geometry.
}}\endinsert}\ni
The usefulness of representing the exact solution in terms of period integrals
on a one complex dimensional geometry rather than in terms of Calabi--Yau 
three-fold periods is quite limited. Note that the Calabi--Yau
representation gives an equally suited description even for the case 
of $G=A_n$. The concept to represent the Weyl transformations generated
by the monodromies of the ADE singularity in terms of an appropriate multiple
cover is less suited for other gauge groups. This is reflected
in the properties of the one complex dimensional geometries which have
been proposed to describe exact solutions from field theory arguments:
for $G=B_n,\ C_n,\ D_n$, one obtains Riemann surfaces with a higher
genus (and therefore a larger number of period integrals) 
then expected from the rank of the gauge group \refs{\othergg,\mw}.
In the case of pure $E_6$ gauge theory the situation is even more complicated
\lw. A systematic expression for the differential $\lambda$ and
a canonical Prym sub-variety of the Jacobian of correct rank have been given in
\mw\ from the connection to integrable systems. 

The increasing difficulties to represent
the field theory solution in terms of period integrals on a Riemann 
surface can be considered as the price for reducing the dimension 
of the geometry from three to one complex dimension. Note that,
differently than the Riemann surfaces which have been suggested
in some cases $G\neq A_n$, the Calabi--Yau
three-fold geometry $S$ has by construction always the correct dimension
of the homology lattice and the periods are in 1-1 correspondence
with the $2\ {\rm rk}(G)$ scalar vev's of the gauge theory. Even more
importantly, {\it the differential form is canonically given by the
unique holomorphic three-form $\Omega$ on $X_3^*$}.
The somehow awkward and unnecessary complications in the case of more general 
gauge groups are the reason why we will concentrate on the $G=A_n$ case in 
the following section, where we derive the
meromorphic one-form and the stable BPS spectrum from reducing
the Calabi--Yau three-fold geometry $S$ to a Riemann surface $\Sigma$
\klmvw. 
It should
be evident that the string theory formulation provides the 
appropriate framework to study the other gauge groups $G$ as well.

\subsec{Meromorphic form and BPS states on the Riemann surface $\Sigma$}
{\it The meromorphic one-form $\lambda$ on $\Sigma$}:
Let us sketch the string theory derivation of the meromorphic one-form
$\lambda$ which enters the effective field theory solution in terms of 
periods on a Riemann surface $\Sigma$ as in \swsii. The starting
point will be  the unique holomorphic three-form $\Omega$ which
enters the definition of the period integrals \tfp.
As an important application we will derive the {\it stable}
BPS spectrum from the string point of view in the next paragraph.

As a concrete example consider the case of pure  $SU(n)$ gauge theory.
The local geometry of the mirror manifold $X_3^*$ is defined
by the zero of a polynomial $W$:
\def\ah#1#2{\hat{a}_{#1}(#2,u_k)}
\eqn\wmg{\eqalign{
W &= \frac{\Lambda^{2n}}{z}+z+2\prod_{i=1}^{n}(x-r_i(u_k)) 
+ w^2+y^2 \cr
&= 2\prod_{i=1}^n(x-\ah{i}{z})+{\ \rm quadratic\  terms} ,
}}
where $u_k,\ k=1,\dots,n-1$ are the
moduli on the Coulomb branch of the $SU(n)$ 
theory and $z$ denotes the coordinate on the base $\IP^1$. The
polynomial $W$ describes an ALE space with 
$A_n$ singularity varying over the compactified $z$ plane. Note that 
at those points $z_{ij}$, where two of the roots of $W$ coincide,
\eqn\vtc{
\ah{i}{z_{ij}}=\ah{j}{z_{ij}} \ ,
}
the ALE space develops a vanishing two-sphere described by the 
local equation $x^{\prime\ 2}+y^2+z^2=0$. More generally, each pair
$(\ah{i}{z},\ah{j}{z})$ defines a two-cycle $C_{ij}$ in the
ALE fiber with a volume that depends on the position $z$ on the base.
There are $\rk(G)=n-1$ independent two-cycles of this type.
The points $z=e_{ij}^\pm$ where \vtc\ holds, 
are at the same time the branch points of the projection to the $z$ plane
and thus the special points associated with monodromies.
Moreover the same two-cycle $C_{ij}$ vanishes
above the two points $e_{ij}^\pm$ on the $z$ plane. 
These pairs of points are
related by the symmetry $z\to \Lambda^{2n}/z$ of the polynomial $W$.
Note that in the second expression in \wmg, the equation for the
Calabi-Yau is well defined but the functions $\ah{i}{z}$ are not
single-valued as functions of $z$; only the product 
$\prod_{i=1}^{n}(x-\ah{i}{z})$
is well defined over $z$. As we move 
around in the $z$-plane, the set of $\ah{i}{z}$ comes back
to itself, but the individual $\ah{i}{z}$ do not necessarily come back
to themselves. In general they are permuted by an element of $S_n$,
the Weyl group of $A_{n-1}$. Since each vanishing cycle is
associated with a pair of $\ah{i}{z}$, the behavior of
the $\ah{i}{z}$ determines the monodromy action on the
vanishing cycles.

The unnormalized unique holomorphic three-form can be represented by
\eqn\omi{
\Omega=\frac{dz}{z}\frac{dx dy}{\p_w W} \ .
}
Integrating $\Omega$ over the two-cycle $C_{ij}$ yields
a one-form $\lambda(z)_{ij}$
\eqn\omii{
\lambda(z)_{ij}=(\ah{i}{z}-\ah{j}{z})\frac{dz}{z} \ .
}
Note that the difference $(\ah{i}{z}-\ah{j}{z})$ is a measure for
the volume of the two-cycle $C_{ij}$ above the point $z$. The one-forms
$\lambda$ are defined on the Riemann surface $\Sigma$ given by 
the vanishing of 
$$
W^{\prime}=\prod_{i=1}^n(x-\ah{i}{z}) \ .
$$
To complete
the integral over a three-cycle in $X_3$ we have to integrate a one-form
$\lambda$
on a path $\gamma(z)$ in the $z$ plane. There are two different types of
such paths which correspond to the image of a three-cycle of $X_3^*$
in the $z$ plane (fig.16):

\item{o} If we transport a two-cycle $C_{ij}$  in the
ALE space along a non-contractible 
closed path on the base, 
we obtain a three-cycle $C_3$ of the topology $S^2\times S^1$.
A D3 brane wrapped on $C_3$ will give rise to a vector multiplet.

\item{o} A three-cycle $\tilde{C}_3$ 
of topology $S^3$ is obtained by starting from a 
point $e^-_{ij}$ on the base where a two-cycle $C_{ij}$  vanishes in 
the ALE space and 
follow a path to the different  point $e_{ij}^+$, where the
same two-cycle $C_{ij}$ vanishes. A D3 brane wrapping 
on $\tilde{C}_3$ gives rise to a matter multiplet. 

Note that we have obtained in this way a map $\phi:\ H_3(X^*_3)\to
H_1(\Sigma)$ which has the property that integrating a one-form
$\lambda$ along $\gamma(z)=\phi(C_3)$ is equivalent to the 
period integral of the three-form $\Omega$ over $C_3$.
\vskip 0.5cm
{\baselineskip=12pt \sl
\goodbreak\midinsert
\centerline{\epsfxsize 3.4truein\epsfbox{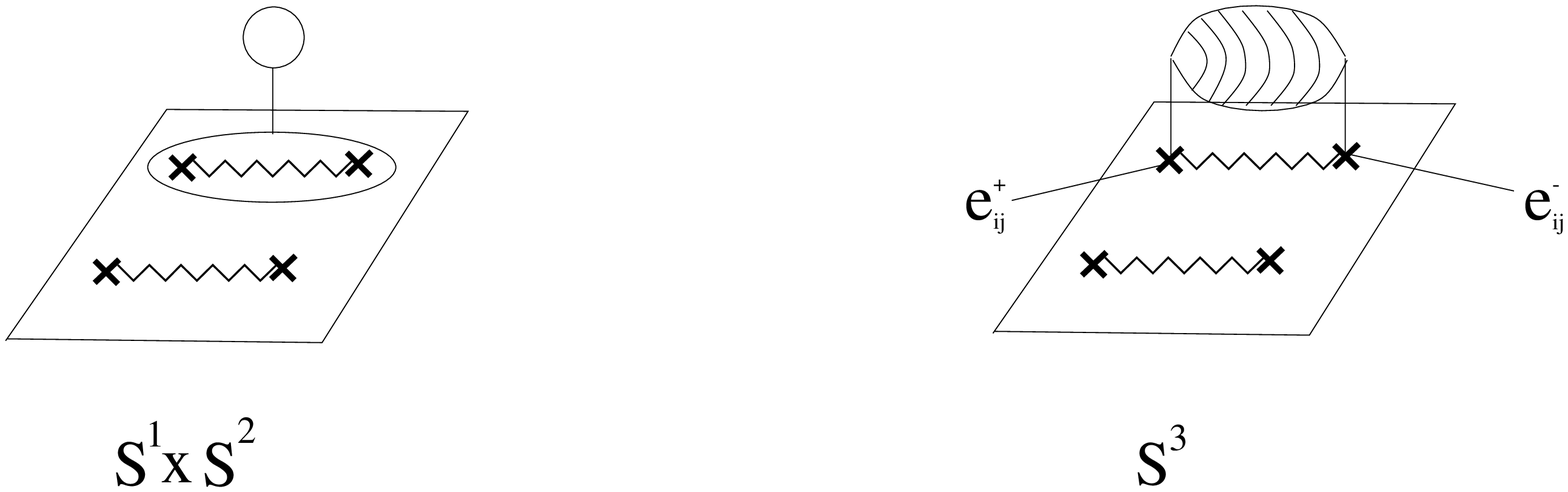}}
\leftskip 1pc\rightskip 1pc \vskip0.3cm
\noindent{\ninepoint  \baselineskip=8pt \centerline{{\bf Fig. 16:}
Three-cycles from two-cycles of the ALE fiber.
}}\endinsert}\ni
\mbr

\nin
{\it Self-dual strings windings and the stable BPS spectrum}:
Let us follow now the BPS states of the type II string, namely
D3 branes wrapped on three-cycles of $S \subset X^*_3$, through the above
reduction to the Riemann surface $\Sigma$. The D3 brane wrapped on 
$C_3$ gives rise to a string stretched on a curve $\gamma(z)$
on $\Sigma$. There are $n-1$ different strings of this kind
corresponding to the $n-1$ independent choices of a two-cycle
$C_{ij}$ of the ALE space. The tension of these strings 
is proportional to the volume of $C_{ij}$, which depends
on the position $z$ on the base. 

The mass of a particle in the four-dimensional theory which
arises from the string with tension $T(z)$,
stretched on a particular path $\gamma(z)$ is
\eqn\bpsmass{
m(\gamma(z))=|
\int_{\gamma(z)}
T(z)\frac{dz}{z}|=|\int_{\gamma(z)}\lambda(z)| \ ,
}
where in the last step
we have used the fact that the mass of the wrapped D3
brane is given by the period integral \tfp. 
Note that the charges of the BPS state
are fixed by the homology class of the path $\gamma(z)$.
Thus the last
expression in eq.\bpsmass\
is precisely the formula obtained in field theory,
$m(q^i_e,q^i_m)=\sqrt{2}|q_e^ia^i+q_m^ia_D^i|$.  Furthermore
note that the string tension $T(z)$ is identified 
with the differences $\ah{i}{z}-\ah{j}{z}$, as expected.
\vskip 0.5cm
{\baselineskip=12pt \sl
\goodbreak\midinsert
\centerline{\epsfxsize 1.6truein\epsfbox{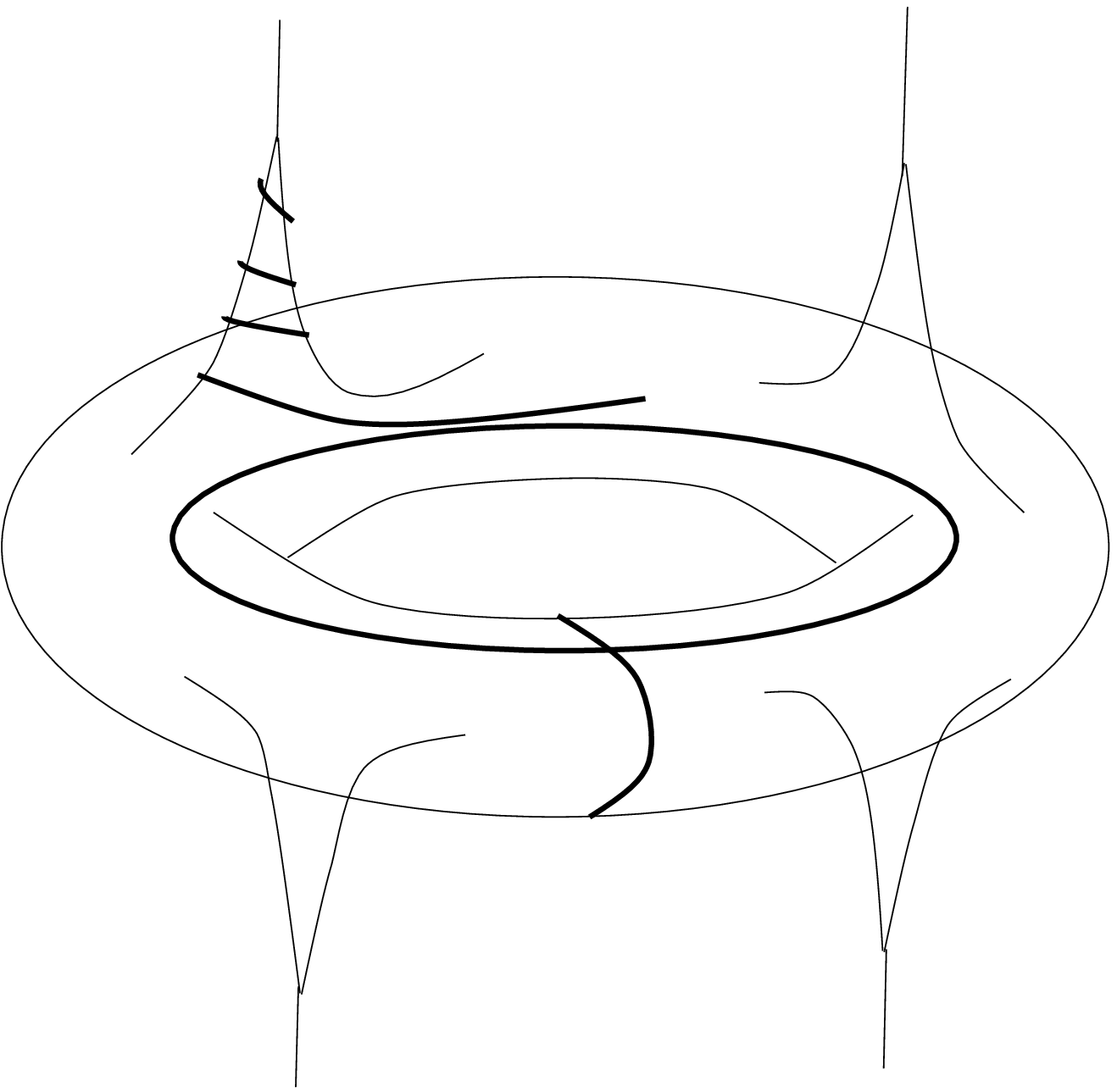}}
\leftskip 1pc\rightskip 1pc \vskip0.3cm
\noindent{\ninepoint  \baselineskip=8pt \centerline{{\bf Fig. 17:}
BPS string windings on $\Sigma$ with non-trivial metric 
$g_{z\bar{z}}\sim\lambda_z\bar{\lambda}_{\bar{z}}$
}}\endinsert}\ni
\nin
Though the BPS mass formula of field theory gives the mass
of an allowed BPS state, it does not say anything about its
existence - a conceptually quite difficult question
in field theory. Here is where the string theory point of view
improves on field theory reasonings. 
What we are interested in is to construct minimal 
volume three-cycles $C_3$ carrying the D3 brane states. For each
point on $\gamma(z)$, there is a two-cycle $C_2$ in the ALE fiber
which is already 
of minimal volume. So to minimize the volume of $C_3$, we
have to minimize the mass of the resulting string stretched
in the $z$ plane. On the latter there is a non-trivial metric arising from
the variation of the string tension with respect to the position 
$z$. The problem is therefore equivalent to consider geodesics in the
metric $g_{z\bar{z}}=\lambda_z\lambda_{\bar{z}}$. The
question of whether a BPS state $\Psi(q_e^i,q_m^i)$ with given 
quantum numbers
exists as a stable state in the spectrum is therefore reduced
to a simple geometric question: {\it the existence of $\Psi(q_e^i,q_m^i)$ is
equivalent to the existence of a primitive geodesic $\gamma(z)$ in the
homology class defined by the quantum numbers $(q_e^i,q_m^i)$}. The 
absence of appropriate geodesics can be visualized as in 
fig.17: for certain quantum numbers, the geodesics run
off to infinity and can not be completed to closed curves in
the appropriate homology class.

\subsec{$N=2$ SYM from the type IIA five-brane}
Let us now describe briefly the third representation
of $N=2$ SYM theory in terms of the world volume theory
of a type IIA five-brane wrapped on a one complex dimensional
Riemann surface $\Sigma$ \klmvw.

Starting from type IIA on a ALE fibration we ended up with
a type IIB mirror geometry $S$, which can be understood
in terms of a ALE space varying over the $z$ plane.
We can now use a different T-duality transformation described in 
ref.\ov, which maps type IIB(A) theory in the neighborhood
of a $A_n$ singularity of the ALE space to type IIA(B) theory on $n$ symmetric
five-branes. In six dimensions, this transformation maps the gauge
symmetry enhancement of the type IIA theory on the $A_{n-1}$ singularity
to type IIB with $n$ symmetric five-branes, which is mapped by 
the $SL(2,{\bf Z})$ of type IIB to $n$ coincident D5-branes with 
enhanced $SU(n)$ gauge symmetry. The wrapped D2-branes are mapped to 
elementary strings stretched between the D5-branes. Similarly, starting
with the type IIB theory on the $A_n$ singularity, one ends
up with type IIA on $n$ symmetric five-branes. The six-dimensional
self-dual string from the D3 brane wrapping on two-cycles corresponds
now to the boundary of a D2-brane ending on the type IIA five-brane.
Note that
we have only to use perturbative string symmetries to reach the
five-brane representation. The same 
configuration has been rederived in ref.\witb\ 
from non-perturbative type IIA/M-theory duality.

Since the space transverse to the ALE space is of
the form $\Sigma\times M_4$, we expect that the five-brane world volume
is compactified on $\Sigma$ after the T-duality map. Let us
see how this works in detail.
The type IIA five-brane
geometry which is T-dual to the type IIB geometry \wmg\ is the following:
the $n$ five-branes are described by the equations
\eqn\xxx{
w=y=0,\qquad x=\ah{i}{z} \ .
}
A collision of five-branes, $\ah{i}{z_{ij}}=\ah{j}{z_{ij}}$ corresponds to 
an $A_1$ singularity of the ALE fiber in the type IIB theory, and similarly for 
the higher singularities. The type IIB string with tension $\sim \ Vol(C_{ij}(z))$
corresponds to the D2-brane stretched between the five-branes at
$x=\ah{i}{z}$ and $x=\ah{j}{z}$.

Now the  
fact that the  $\ah{i}{z}$ vary holomorphically with  $z$ implies that
the several world volumes of the $n$ five-branes located
at the $\ah{i}{z}$ are joined together and combine effectively 
to a {\it single} five-brane given by
$$
\Sigma \times M_4\ ,$$
where the $M_4$ is the four-dimensional Minkowski space-time,

The resulting four-dimensional gauge theory is obtained from
dimensional reduction of the six dimensional world volume theory
on $\Sigma$. In six dimensions, the spectrum consists of a self-dual
two-form $B$ and five real scalars. On compactification
on $\Sigma$, the world volume theory is twisted \vafaiii;
from the two-form we get $h^{1,0}={\rm rk}(G)$ vector bosons.
Moreover two of the five scalars become one-forms on $\Sigma$ upon
twisting and give rise to $2\ {\rm rk}(G)$ scalars which combine
with the gauge bosons to complete ${\rm rk}(G)$ four-dimensional 
vector multiplets. The remaining three scalars are unaffected by the twist
and do not give rise to four-dimensional fields due to the absence of
normalizable zero modes on the non-compact Riemann surface $\Sigma$ 
of infinite volume. Note that the vev's of
the scalars in the vector multiplets, which correspond to the volumes
of the three-cycles in the type IIB theory, can be identified with the
Seiberg--Witten differentials $\lambda$:
\eqn\las{
\langle \phi \rangle = \lambda \ .
}
Note that the variation of $\lambda$
with respect to the zero mode of $\phi$ yields harmonic
one-forms on $\Sigma$ \sw.  The latter give rise to the 
dynamical scalar degrees of freedom of the vector multiplets,
in agreement with the identification made in eq.\las.

In summary, we have reached a T-dual representation of the $N=2$
SYM theory in terms of a type IIA five-brane world volume theory 
on $\Sigma \times M_4$ embedded in an eight-dimensional space
$(x,z,M_4)$. The metric on the $x$-plane is the flat
metric and on the $z$-plane the metric is cylindrical, given by
$|dz/z|^2$. The BPS states correspond to two-branes ending on the
five-branes with boundaries wrapped on non-trivial cycles of $\Sigma$
(fig.18).
Moreover the  tension is given by $|dx dz/z|$.
\vskip 0.5cm
{\baselineskip=12pt \sl
\goodbreak\midinsert
\centerline{\epsfxsize 2truein\epsfbox{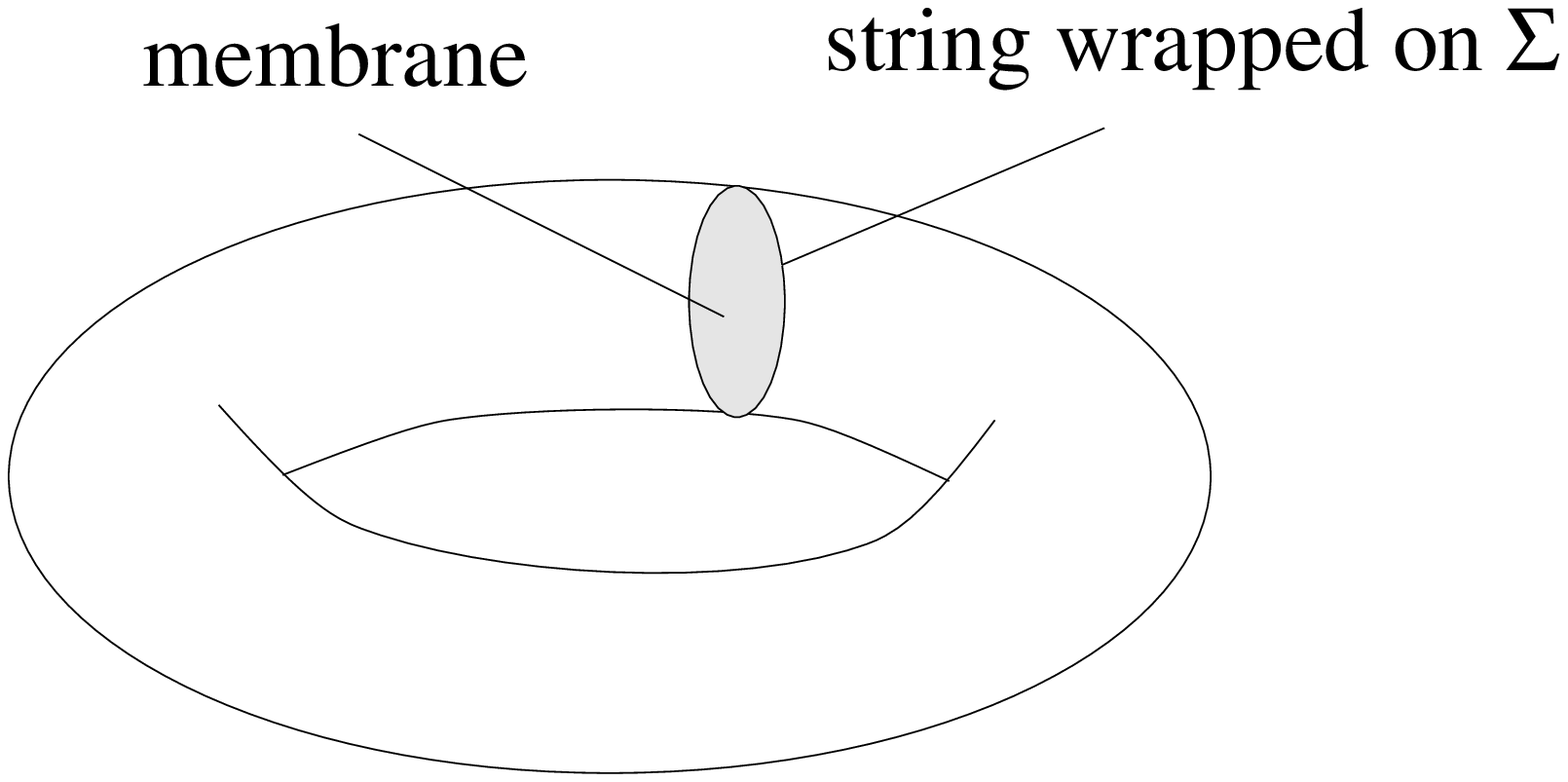}}
\leftskip 1pc\rightskip 1pc \vskip0.3cm
\noindent{\ninepoint  \baselineskip=8pt \centerline{{\bf Fig. 18:}
Membrane BPS states ending on the type IIA five-brane.
}}\endinsert}\ni
Though the five-brane picture is appealing, there are some good
reasons to concentrate on the T-dual Calabi--Yau 
representation\foot{The situation is somehow different in the case of $N=1$
theories, to which the five-brane picture has been extended to 
some extent \neo, 
while a geometric analysis is still lacking.}.
Firstly the special geometry of the Calabi--Yau moduli space
represents a strong mathematical framework to determine 
the exact effective action and physical states: in particular the 
metric on the moduli space as given by the unique holomorphic three-form,
the period integrals as well as a treatment of BPS states as
described above. Importantly, there is no restriction on the gauge
group $G$, differently then in the five-brane
picture.
Also, we get for free the exact effective action describing
the coupling of the gauge system to gravity, a question which has not been
addressed so far in the brane language.

\newsec{Outlook}
In short, we have seen that type II string theory can be considered as the
natural underlying structure of the Seiberg--Witten theory. It gives a 
concrete physical meaning to the Riemann surface $\Sigma$ and 
provides a ratio for the appearance of three-dimensional 
Calabi--Yau geometries in the exact solution of SYM theories with
general gauge groups. It gives a unified description of all 
magnetic and electric 
BPS states in terms of 
self-dual strings which wind on the non-trivial, geodesic homology
cycles of $\Sigma$. 

Even more, the string theory approach provides a powerful
tool to generate and study a large class of $N=2$ theories from
a systematic study of geometrical Calabi--Yau singularities
together with D-brane technology. This class includes gauge
theories in $d\leq 6$ with arbitrary gauge groups, interacting
conformal field theories and more exotic theories involving
non-critical strings. A systematic study has been started 
in \kmv\ for the subclass of theories with only $A$ type 
fiber singularities; some interesting results are

\item{o} The exact solution of all asymptotic free
$N=2$ SYM theories with gauge group 
$\prod_i SU(n_i)$ with bi-fundamental and fundamental matter.
\item{o} The classification of superconformal theories in the above
class in terms of affine ADE Dynkin diagrams.
\item{o} The relation of the gauge coupling space of these theories
to the moduli
of flat ADE connections on a torus.
\item{o} The $S$-duality groups of all these theories in terms
of the fundamental group of flat ADE connections on the torus.
\item{o} The interpretation of the $S$ duality groups as the
effective duality group acting on $\tau_{\rm eff}(a_i)$
of a different gauge theory.
\item{o} A new duality of $d\leq 5$ dimensional $N=2$ theories, e.g. 
relating a $SU(n)^m$ theory to a $SU(m+1)^{n-1}$ theory.
\item{o} New exotic $N=2$ theories containing the coupling constants
of a SYM theory as dynamical fields.

\nin
It will be interesting to complete this program for other
fiber singularities
and to analyze the new physics of those
theories, which do not correspond to known Lagrangian
field theories.
\mbr

\nin
{\it Acknowledgments:} The research of the author was supported by
NSF grants PHY-95-13835 and PHY-94-07194.

\listrefs

\end